\documentclass[useAMS,aas_macros,usenatbib]{mn2e}
\usepackage{epsfig,subfigure}
\usepackage{latexsym,amssymb}

\newcommand{\sgn} {\ensuremath{\mathcal{S}}}
\newcommand{\Msun}{\ensuremath{M_\odot}}

\begin{document}

\title[Fossil Gas and Precursors of BBH Mergers]{Fossil Gas and the Electromagnetic Precursor of Supermassive Binary Black Hole Mergers}
\author[Chang, Strubbe, Menou,\& Quataert]{Philip
  Chang$^{1,2}$\thanks{E-mail: pchang@astro.berkeley.edu (PC); linda@astro.berkeley.edu (LES); kristen@astro.columbia.edu (KM); eliot@astro.berkeley.edu (EQ)}, 
  Linda E. Strubbe$^{1}$\footnotemark[1], 
  Kristen Menou$^{3}$\footnotemark[1], \& 
  Eliot Quataert$^1$\footnotemark[1]
  \\$^{1}$ Department of Astronomy and
  Theoretical Astrophysics Center, 601 Campbell Hall, University of
  California, Berkeley, CA 94720
  \\ $^{2}$ Miller Institute for Basic Research
  \\ $^{3}$ Department of Astronomy, Pupin Hall, Columbia University, 550 West 120th Street, New York, NY 10027
} 


\maketitle

\begin{abstract}

  Using a one-dimensional height integrated model, we calculate the
  evolution of an unequal mass binary black hole with a coplanar gas
  disc that contains a gap due to the presence of the secondary black
  hole.  Viscous evolution of the outer circumbinary disc initially
  hardens the binary, while the inner disc drains onto the primary
  (central) black hole.  As long as the inner disc remains cool and
  thin at low $\dot{M}_{\rm ext}$ (rather than becoming hot and geometrically
  thick), the mass of the inner disc reaches an asymptotic mass
  typically $\sim 10^{-3}-10^{-4}\Msun$.  Once the semimajor axis shrinks below a
  critical value, angular momentum losses from gravitational waves
  dominate over viscous transport in hardening the binary. 
  The inner disc then no longer responds viscously to the inspiraling black
  holes.  Instead, tidal interactions with the secondary rapidly drive the inner disc into the
  primary. Tidal and viscous dissipation in the inner disc lead to a
  late time brightening in luminosity, $L\propto t_{\rm minus}^{5/4}$, where $t_{\rm minus}$ is the time
  prior to the final merger.  This late time brightening peaks $\sim 1$ day prior to the
  final merger at $\sim 0.1 L_{\rm Edd}$. This behavior is relatively robust
  because of self regulation in the coupled viscous-gravitational
  evolution of such binary systems.  It constitutes a unique electromagnetic 
  signature of a binary supermassive black hole merger and may allow the host 
  galaxy to be identified if used in conjunction with the Laser Interferometric 
  Space Antenna (LISA) localization.
\end{abstract}


\section{Introduction}\label{sec:intro}

There is now strong evidence that supermassive black holes (SMBHs)
inhabit the spheroids of most galaxies
\citep{Magorrian1998,Tremaine2002}.  Galaxy mergers, which are a
natural consequence of hierarchical assembly of galaxies, should cause
the SMBHs in each galaxy to form a binary in the newly 
assembled host galaxy \citep{Begelman1980}.  These binary black holes
(BBHs) are believed to coalesce due to an uncertain combination of
dynamical friction, three body interactions, interaction with viscous
circumbinary discs, and gravitational wave (GW) losses.  Observations
of the GW that these systems produce is 
one of the primary scientific objectives of the proposed Laser
Interferometer Space Antenna (LISA).  In addition to their usefulness
as probes of general relativity in the strong 
field regime, GWs can also be used to measure the luminosity distance
to high accuracy, i.e., $\delta D_{\rm L}/D_{\rm L}$ at the percent
level in many cases \citep{Hughes2002,Holz2005}.

The high accuracy that these ``standard sirens'' offer would allow these
GW sources to constrain cosmological models if the host galaxy can be
identified, so that a source redshift can be measured.  By
itself, LISA observations will provide sky localization errors that
are typically $\sim$ few $\times 10^{-1}$ deg$^2$ \citep[see for
  instance][]{Holz2005,Kocsis2007,Kocsis2008a,Lang2008,Arun2008},
in which there may be $\sim 10^4$ galaxies.\footnote{The 3-D
  localization of LISA would give $\sim$ 10 candidate galaxies per
  arcmin$^{-2}$ \citep{Holz2005,Kocsis2006}.}  Hence, much recent work
has focused on finding an unambiguous electromagnetic counterpart to a
BBH merger \citep{Schutz1986} to help identify its host galaxy. In
addition, the known BH masses and spins from GWs, an observed
electromagnetic counterpart would also provide strong constraints on
models of accretion discs and quasars/AGNs.

At present, most studies of these electromagnetic counterparts to
BBH mergers have focused on the post-merger phase.
At the time of merger, a fraction of the rest mass of the two BHs is
radiated as GW and the merged BH receives a kick between
$100-4000\,{\rm km\,s^{-1}}$
\citep{Bekenstein1973,Favata2004,Baker2006,Herrmann2007,Gonzalez2007,Boyle2008}.
If a circumbinary gas disc surrounded the two BHs (pre-merger), 
the sudden decrease in BH mass and the kick provided by the BH
may produce an observable electromagnetic signature
\citep{Bode2007,Lippai2008,Schnittman2008,Shields2008,Oneill2008,Bode2009}.
On a longer timescale, the circumbinary disc would accrete
into the newly merged BH, leading to AGN activity \citep{Milos2005} or
perhaps an offset quasar \citep{Madau2004}.  

There are 
potential difficulties associated with identifying these
electromagnetic signatures of BBH mergers.  The signatures associated
with kicks are likely to be weak and difficult to distinguish from AGN
variability, 
if they are a few percent of the background disc emission
\citep{Schnittman2008,Kocsis2008b,Oneill2008}.  Delayed AGN and offset
AGN activity, while significantly more promising, might have a
substantial delay \citep{Milos2005,Loeb2007}.

The subject of this work is to explore electromagnetic emission
associated with the {\it pre-merger} phase.  Various authors
\citep{Milos2005,Armitage2005,MacFadyen2008} have studied the physics
of such viscously driven BBH mergers.  These studies suggest that the
interaction between the disc and the BBHs can induce a small
eccentricity in the disc and the binary
\citep{Armitage2005,MacFadyen2008,Cuadra2009}.  In addition, some of
this gas may form accretion discs around the two BHs leading
to binary AGN activity \citep{Dotti2007}.

In this paper, we build on the work of \cite{Armitage2002} and show
that although the inner disc drains when the binary is well separated,
a small amount of remnant or fossil gas can remain around the primary
BH, giving rise to a distinct electromagnetic
signature.  The amount of fossil gas is very small ($\sim
10^{-3}\Msun$), but even this tiny amount of gas is sufficient to
drive the luminosity of the binary to $\sim 0.1\,L_{\rm Edd}$ one day
before merger.

The plan of this work is as follows.  In \S\ref{sec:basic picture}, we
summarize the basic scenario. We then present the relevant physics in
\S\ref{sec: basic equations}. In \S\ref{sec:results}, we argue that a
small amount of fossil gas could exist in the inner disc between the
primary and secondary BHs and that this fossil gas would be tidally
driven into the primary on a very short timescale near the final
merger, leading to a bright electromagnetic precursor.  Finally in
\S\ref{sec:conclusions}, we discuss some of the caveats of this model and
directions for future work.

\section{Basic Picture}\label{sec:basic picture}

\begin{figure*}
\begin{center}\includegraphics[width=10cm]{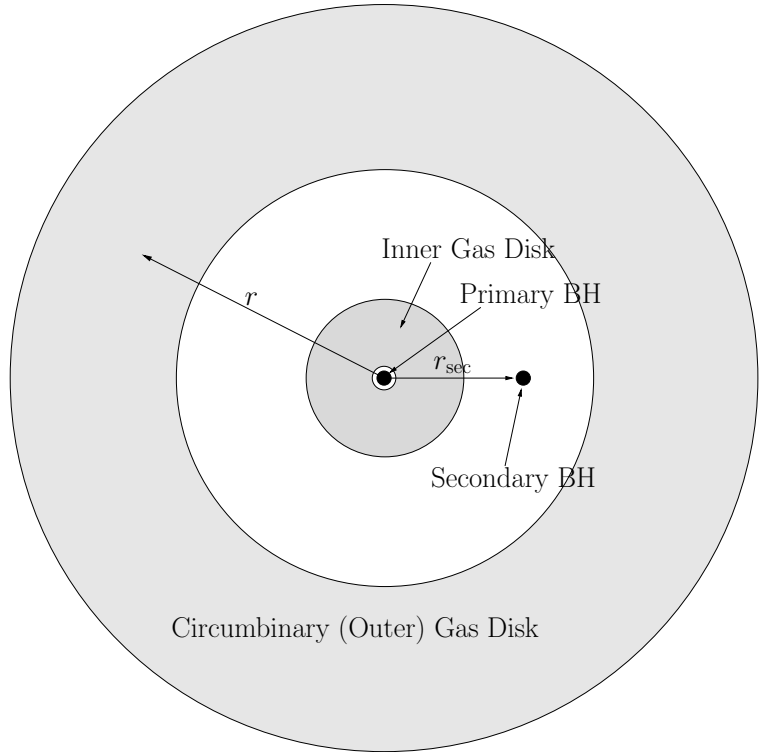}\end{center}
\caption{
Schematic diagram of the model considered in this paper.  An
  accretion disc and a secondary (coplanar) BH with semimajor axis
  $r_{\rm sec}$ surrounds a central BH.  The secondary BH opens up a
  gap in the disc, dividing the disc into an inner disc and an outer
  disc.  The inner disc steadily drains into the central BH, while the
  outer disc viscously spreads and tidally interacts with the secondary.
  The binary's semimajor axis decreases due to tidal interaction with
  the viscous disc until $r_{\rm sec}<r_{\rm GW}$, at which point GW
  emission becomes the primary mechanism by which angular momentum is
  lost.  For $r_{\rm sec} \ll r_{\rm GW}$, the {\it inner} disc can no
  longer viscously respond on the merger timescale and is driven into
  the primary by tidal interactions with the secondary.
}
\label{fig:cartoon}
\end{figure*}

We illustrate the basic scenario schematically in Figure
\ref{fig:cartoon}.  A secondary BH with mass $M_{\rm sec}$ and a
coplanar accretion disc orbits a more massive primary BH at a radius
$r_{\rm sec}$.
We expect the disc to be aligned to the orbital plane of the two BHs because of inclination damping. \cite{Ivanov1999} showed that in
the case of a BH orbiting a more massive BH with an associated
accretion disc, the inner accretion disc aligns with the orbit of the
binary independent of the viscosity.  As a result,
the accretion disc is truncated into two parts: an inner accretion
disc which surrounds the primary and a circumbinary disc which
surrounds the binary.  Flow between the inner disc and outer disc is
prevented by the presence of the secondary BH at least in one
dimension.  The inner accretion disc has an inner radius, which is
typically the innermost stable circular orbit (ISCO), and an outer
radius where it is truncated due to tidal interactions with the
secondary BH.  Similarly, the outer disc is also truncated.

A similar model has previously been studied by \cite{Armitage2002}.
However, we will include several improvements in our treatment.  First
we include more realistic physics for the evolution of the viscous
disc, i.e., we use a self-consistent alpha disc instead of a fixed
viscosity law.  This significantly changes how the disc drains and we
believe that it provides a more realistic estimate of the luminosity
evolution of the disc prior to merger.  Secondly, we focus on the 
luminosity of the late time electromagnetic precursor and 
 its simple shape, which might aid its identification.

\section{Basic Equations}\label{sec: basic equations}

We now present the governing equations for our model.  These are
standard in the literature \citep{Lin1979a,
  Lin1979b,Lin1986,Hourigan1984,Ward1989,Rafikov2002,Chang2008,VanDeVen2008},
but we state them here for completeness.  We adopt the notation of
\cite{Chang2008} and \cite{VanDeVen2008}.  We also include the effect
of angular momentum losses due to GWs, which become important at small
binary separation.

The evolution of a viscous disc under the influence of an external
torque is given by equations for continuity and angular momentum
conservation.\footnote{For simplicity, we do not make a distinction
  between the center of mass of the binary and the location of the
  primary BH in our analysis, even though the secondary is not much
  less massive than the primary in some of our models.}  The equation of
continuity is \citep{Frank2002}
\begin{equation}\label{eq:continuity}
  \frac {\partial \Sigma} {\partial t} + \frac 1 r \frac {\partial
    (r\Sigma v_r)}{\partial r} = 0,
\end{equation}
where $r$ is the radial coordinate and $v_r$ is the radial component
of the velocity.  The equation of angular momentum conservation is
\citep{Pringle1981,Rafikov2002}:
\begin{equation}\label{eq:full angular momentum}
  \frac{\partial (\Sigma r^2\Omega)}{\partial t} + \frac 1 r
  \frac{\partial (r v_r \Sigma r^2 \Omega)} {\partial r} = -\frac 1
  {2\pi r}\left(\frac {\partial T_{\rm visc}}{\partial r} - \frac
    {\partial T_{\rm d}}{\partial r}\right) ,
\end{equation}
where $\Omega = \sqrt{GM_{\rm BH}(r)/r^3}$ is the orbital frequency,
$M_{\rm BH}$ is the mass of the primary BH, and
\begin{equation}\label{eq:viscous torque}
T_{\rm visc} = -2\pi r^3\nu\Sigma\frac {\partial \Omega}{\partial r},
\end{equation}
is the viscous torque, where $\nu$ is the viscosity. We have adopted
the standard $\alpha$ prescription for the viscosity
\citep{Shakura1973}:
\begin{equation}
  \nu = \frac {2\alpha P_{\rm g}}{3\Omega\rho},
\end{equation}
with proportionality to $P_{\rm g}$, the midplane gas pressure.  We
chose to rely on the so-called ``$\beta$ model'', i.e., stress
proportional to gas pressure as opposed to total pressure, to ensure
viscous and thermal stability in our 1-d model.  For most of our
calculations, we use $\alpha = 0.1$.\footnote{The exception
is the suite of calculations presented in Table 1, where we have
take $\alpha = 0.01$ and $\alpha = 0.1$} A simple
$\alpha$-prescription applied to radiation pressure dominated discs is
both viscously and thermally unstable \citep{Lightman1974,Piran1978}.
Recent work by \cite{Hirose2009} suggests that the
thermal instability is avoided in radiation pressure dominated
situations because stress fluctuations lead the associated pressure
fluctuations. However, the viscous instability of radiation pressure dominated
discs might remain.  Adopting the ``$\beta$ model'' for the viscous
stress will only change the evolution at high $\dot{M}_{\rm ext}$ and small
radius and does not affect our key results: 
once the binary is at small separation, the discs' evolution is dominated by
GW losses
and not by
internal viscosity.


Using equation (\ref{eq:continuity}), we simplify (\ref{eq:full
  angular momentum}) to be
\begin{equation}\label{eq:angular momentum}
\Sigma\frac{\partial (r^2\Omega)}{\partial t} + v_r \Sigma
\frac{\partial (r^2 \Omega)} {\partial r} = -\frac 1 {2\pi
  r}\left(\frac {\partial T_{\rm visc}}{\partial r} - \frac {\partial
    T_{\rm d}}{\partial r}\right). 
\end{equation} 
Solving for $v_r$ in
equation (\ref{eq:angular momentum}) and plugging it into
(\ref{eq:continuity}), we find
\begin{equation}\label{eq:time dependent sigma}
\frac{\partial \Sigma}{\partial t} = \frac 1 {2\pi r} \frac {\partial}
{\partial r} \left(\frac{\partial (r^2\Omega)}{\partial r}\right)^{-1}
\left[\frac {\partial}{\partial r}\left( T_{\rm visc} - T_{\rm
d}\right)\right].
\end{equation}

The orbiting secondary BH exerts a torque density on the disc,
$\partial T_{\rm d}/\partial r$, of the form
\citep{Goldreich1980,Lin1979a,Ward1997,Armitage2002,Chang2008}:
\begin{equation}\label{eq:tidal torque}
\frac {\partial T_{\rm d}}{\partial r} \approx \frac 4 9 f{\rm
sgn}\left(r-r_{\rm sec}\right) \frac {G^2 M_{\rm sec}^2 \Sigma
r}{\Omega^2\left({r - r_{\rm sec}}\right)^{4}}, 
\end{equation}
where $f$ is a constant which is $\approx 0.1$ following
\cite{Armitage2002}. 

Plugging equations (\ref{eq:viscous torque}) and (\ref{eq:tidal
  torque}) into (\ref{eq:time dependent sigma}), we find
\begin{eqnarray}
\frac{\partial \Sigma}{\partial t} &=& \frac 1 {r} \frac {\partial}
{\partial r} \left[3 r^{1/2} \frac {\partial}{\partial r}\left( \nu \Sigma r^{1/2}\right) \right.\nonumber\\
&&\left. - \frac {4} {9\pi} f{\rm
sgn}\left(r-r_{\rm sec}\right)\Omega r^2 q^2 \Sigma
\left(\frac{r}{\left({r - r_{\rm sec}}\right)}\right)^4\right],\label{eq:master1}
\end{eqnarray}
where $q = M_{\rm sec}/M_{\rm BH}$ is the mass ratio between the
secondary and primary BHs.

The radial motion of the secondary BH, assuming a circular orbit, is
governed by tidal torquing of the disc and the torque due to GWs:
\begin{eqnarray}
  \frac 1 2 M_{\rm sec} \Omega_{\rm sec} r_{\rm sec}\frac {\partial r_{\rm sec}}
 {\partial t} &=& \frac 4 9 f G M_{\rm BH} q^2
  \int \sgn \left(\frac r {r -
      r_{\rm sec}}\right)^4 \Sigma dr \nonumber\\ &&-
    T_{\rm GW}.\label{eq:rdot}
\end{eqnarray}
where $\sgn = {\rm sgn}\left(r-r_{\rm sec}\right)$ is the sign of the torque, and $T_{\rm GW}$ is the torque exerted by gravitational wave losses \citep{Peters1964},
\begin{equation}
T_{\rm GW} = \frac{\sqrt{8}}{5}M_{\rm BH} c^2\left(\frac {r_{\rm g}}{r_{\rm sec}}\right)^{7/2}\left(\frac{M_{\rm sec}}{M_{\rm BH}}\right)^2\sqrt{\frac{M_{\rm sec} + M_{\rm BH}}{M_{\rm BH}}},
\end{equation}
where $r_{\rm g} = 2GM_{\rm BH}/c^2$ is the Schwarzschild radius.

Finally we assume vertical energy balance at every radius between
heating and radiative cooling, i.e. $F = D(r)$, where $F$ is the local
radiative flux and $D$ is the local dissipation rate.  In a one-zone
model, the local radiative flux is
\begin{equation}
F = \frac 4 3 \frac {\sigma T^4}{\kappa \Sigma},  
\end{equation}
where $T$ is the central temperature and $\kappa$ is the opacity,
which we assume to be constant
(Thomson) for simplicity. More
detailed models would include variations in opacity with radius in the
disc. We assume that the disc emits as a blackbody.  The heating is
given by a combination of viscous dissipation and tidal dissipation.
Viscous dissipation is given by the standard formula \citep{Frank2002}
\begin{equation}
D_{\rm visc} = \frac 9 8 \nu \Sigma \frac {GM_{\rm BH}}{r^3},
\end{equation}
where $D_{\rm visc}$ is the local viscous dissipation rate per unit
area.  

We now consider the energy dissipation from tidal torques.  We assume that the 
torque on the disc raised by the satellite, $T_{\rm d}$, is mediated by the 
excitation of spiral density waves and locally damped.
The timescale associated with this tidal torque is the angular momentum of the 
satellite, $L_{\rm sec} = M_{\rm sec} \Omega_{\rm sec} r_{\rm sec}^2$, divided 
by the torque or $t_{\rm tide} =  L_{\rm sec}/T_{\rm d}$.  The power associated 
with such a torque is then 
\begin{equation}
\dot{E}_{\rm tide} = \frac {E_{\rm sec}}{t_{\rm tide}} = T_{\rm d} \Omega_{\rm sec}, 
\end{equation}
where $E_{\rm sec} = GM_{\rm BH}M_{\rm sec}/r_{\rm sec}$ is the orbital energy of the 
secondary.
Because we assume local damping of spiral density waves, the corresponding binding 
energy liberated locally (assuming circular orbits) is in proportion to the spiral density waves 
which mediate this interaction.  Hence the local tidal dissipation rate is given by
\begin{equation}
D_{\rm tide} = \frac 1 {2\pi r} \dot{E}_{\rm tide} \frac {|dT_{\rm
    d}/dr|} {\int dr |dT_{\rm d}/dr|}.
\end{equation}

As we will show, these tidal interactions are especially important at late times 
when GW losses drive the secondary inward on a circular orbit. A tiny 
fraction ($M_{\rm d, in}/M_{\rm sec} \sim 10^{-10}$, where $M_{\rm d,in} 
\equiv 2\pi \int_0^{r_{\rm sec}} \Sigma r^2dr$ is the mass of the inner disc) 
of the angular momentum losses is offset by a gain from the tidal interaction 
with the inner disc. The tidal-GW evolution allows this fraction 
of the prodigious gravitational luminosity $L_{\rm GW} < 10^{55}\,{\rm ergs\,s}^{-1}$,\footnote{The 
GW luminosity can maximally be $c^5/G \sim 3\times 10^{59}\,{\rm ergs\,s}^{-1}$,
but the GW luminosity when the disc is still present is closer to $\sim 10^{55}\,{\rm ergs\,s}^{-1}$}
to be released electromagnetically by heating the inner disc gas, which is 
enough to power the inner accretion disc close to its Eddington luminosity.

\section{Results}\label{sec:results}

We solve equations (\ref{eq:master1}) and (\ref{eq:rdot}) using
standard explicit finite-difference methods \citep{Press1992}.  
We choose 200 grid points logarithmically spaced between $r =
r_{\rm ISCO} = 3 r_{\rm g}$ (i.e., the ISCO) and $10^5 r_{\rm g}$. We
set a zero-torque boundary condition at the inner radius and impose an
outer boundary condition such that there is a constant external
feeding rate, $\dot{M}_{\rm ext}$.  As a test, we initially solve these
equations for no satellite and with inflow boundary conditions and
find that we recover the steady state $\alpha$-disc solution with an
error of $\lesssim 1\%$.

The first two cases we consider are a $10^7\Msun$ BH with a $q=0.1$
and a $q=0.3$ secondary, respectively.  We start the secondary
at an initial radius of $r_{\rm s,0} = 10^4\,r_{\rm g}$ ($\approx
0.01$ pc).  We begin with a low mass ($10^3\Msun$) disc which extends
from the ISCO to $r_{\rm outer} = 10^5
r_{\rm g}$.  We clear a region around the secondary's initial radius,
$0.5 r_{\rm s,0} < r < 2 r_{\rm s,0} $, to model the initial clearing
of a gap around the secondary BH.  At the outer edge of the grid, we
consider an outer $\dot{M}_{\rm ext}$ of either $\dot{M}_{\rm ext} =
0.1\,\Msun\,{\rm yr}^{-1}$ or $0.01\,\Msun\,{\rm yr}^{-1}$.

We choose a low initial disc mass ($10^3\,\Msun$, which corresponds to
a very low accretion rate in steady-state) so that our models can be
integrated on a reasonable timescale.  The reason is that the
Courant condition, which is set by the viscous time of the disc near
the ISCO, limits the timestep to very low values unless the disc has a
very low mass. Fortunately, our results do not depend on this initial
condition. The mass of the inner disc, $M_{\rm d,in}$, 
approaches an asymptotic value which is
independent of its initial value.  With larger initial disc masses,
the inner disc would have initially drained on a faster time scale,
but the final disc mass would remain the same.

\begin{figure}
\begin{center}\includegraphics[width=8cm]{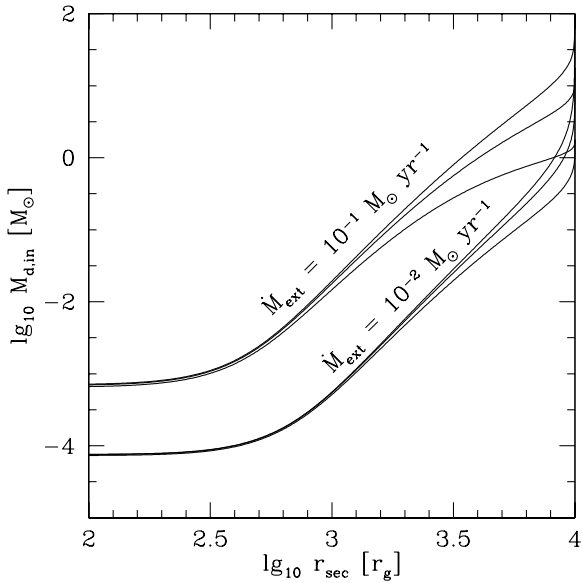}\end{center}
\caption{Evolution of the disc mass inside the secondary's orbital
  radius, $r_{\rm sec}$, for a $10^7\,M_{\odot}$ primary and
  $10^6\,M_{\odot}$ secondary and external mass accretion rates of
  $10^{-1}\,M_{\odot}yr^{-1}$ ($\dot{M}_{\rm ext} \approx \dot{M}_{\rm Edd}$;
  top set of curves), and $10^{-2}\,M_{\odot}yr^{-1}$ ($\dot{M}_{\rm ext}
  \approx 0.1\dot{M}_{\rm Edd}$; bottom set of curves). Three
  scenarios in each case, with initial disc masses of $10^4$, $10^3$,
  and $10^2\,\Msun$, are represented by the curves from top to bottom.
  Note that the asymptotic disc mass is independent of the initial
  disc mass.}
\label{fig:diff_initial}
\end{figure}

We illustrate this point in Figure \ref{fig:diff_initial} where we
show models with different disc masses of $10^2$, $10^3$,
and $10^4\Msun$ for two different mass accretion rates: $\dot{M}_{\rm ext} =
10^{-1}\,M_{\odot}yr^{-1}$ (top set of curves) and $\dot{M}_{\rm ext} =
10^{-2}\,M_{\odot}yr^{-1}$ (bottom set of curves).  Note that the mass
of the inner disc approaches an asymptotic value that is independent
of its initial value.  The reason for this is that the inner disc
drains until its viscous time is comparable to the merger
time.  
At late times, however, when the evolution of the system is
dominated by GWs, the inner disc can no longer viscously drain on the
timescale of the merger.  Hence, the inner disc mass is essentially
frozen.  The frozen mass of the inner disc at late times is given by the
mass of the inner disc when the system transitions from
viscosity-dominated evolution to GW-dominated evolution, which occurs
typically at $r_{\rm GW} \sim 500 r_{\rm g}$ \citep[ also see
  \S4.2]{Armitage2002,Haiman2008}.
In the remainder of this work, we 
use this fact and start with a
low disc mass to save computational cost.

\subsection{Viscous Evolution of the BBH and the Outer Disk}

\begin{figure*}
\begin{center}
\subfigure[]{
\includegraphics[width=.45\textwidth]{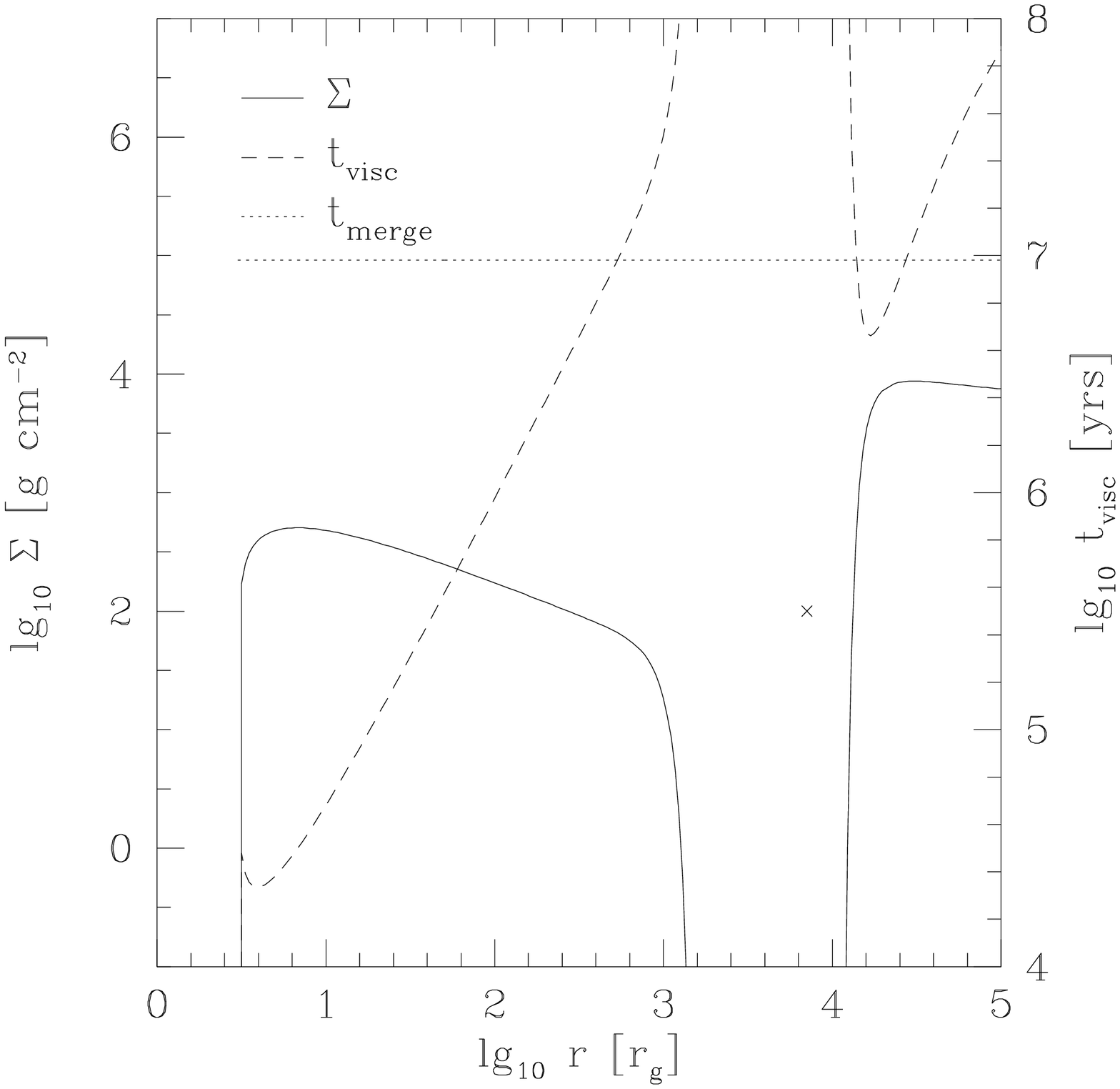}}
\subfigure[]{
\includegraphics[width=.45\textwidth]{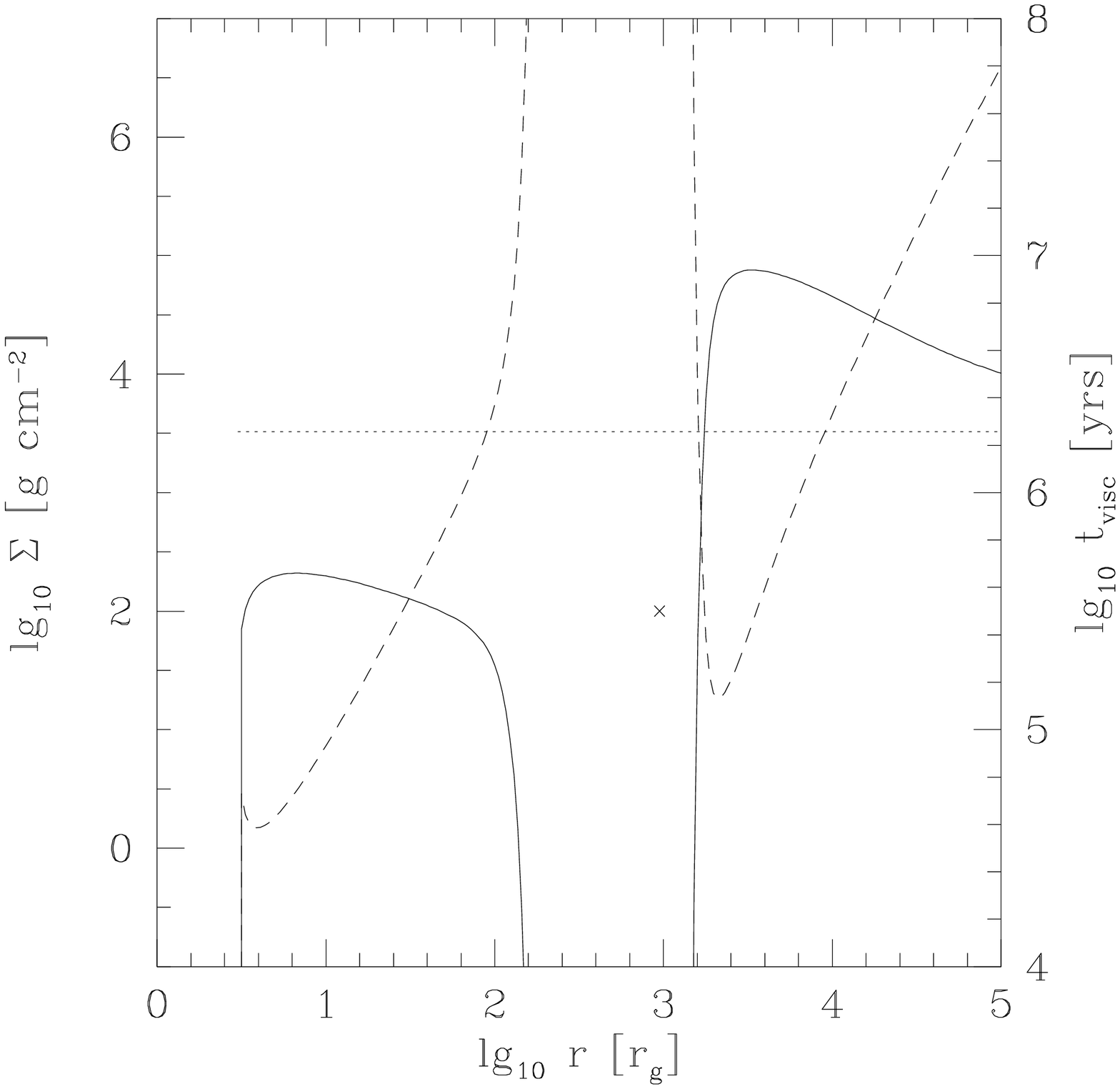}}\\
\subfigure[]{
\includegraphics[width=.45\textwidth]{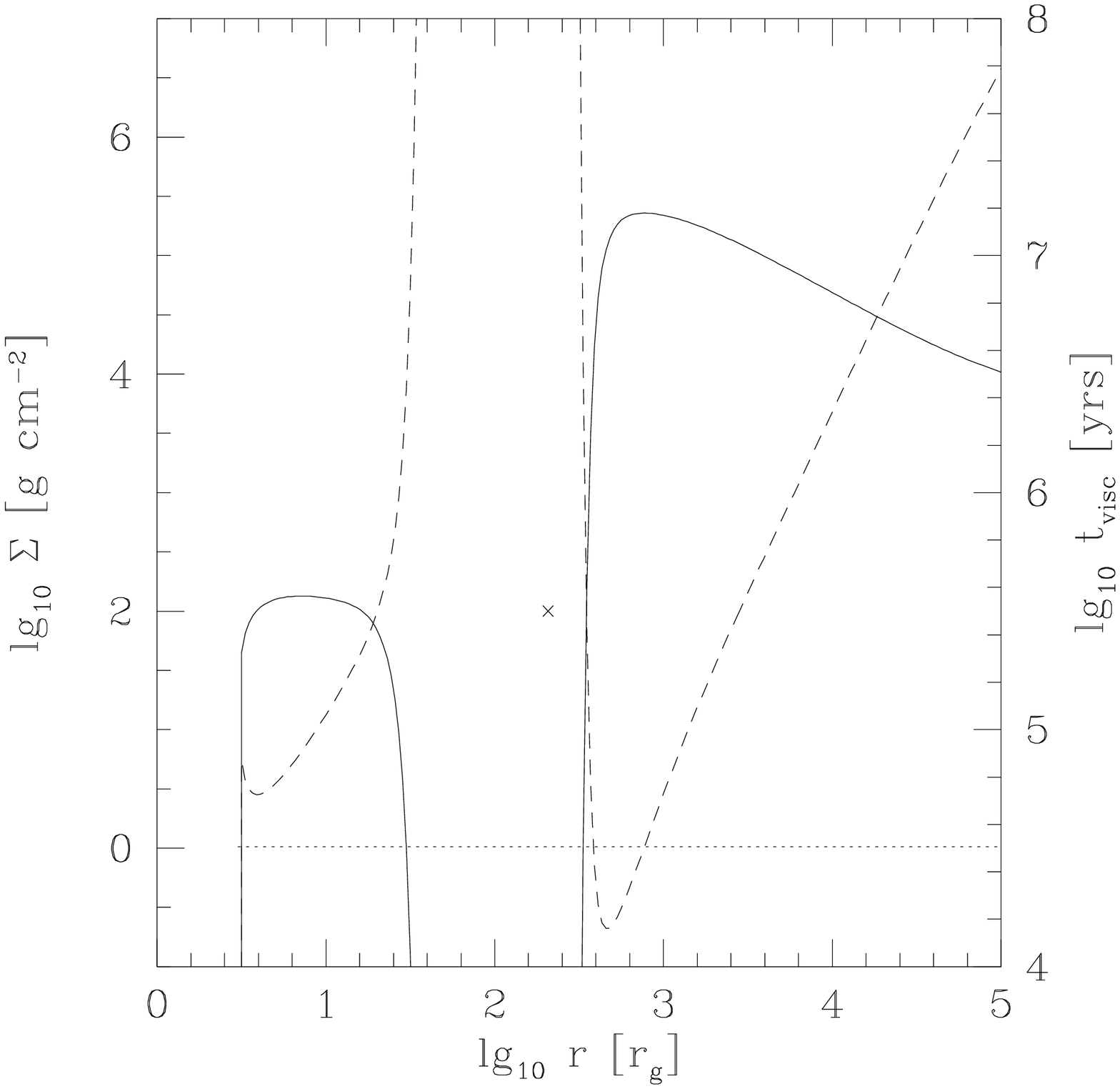}}
\subfigure[]{
\includegraphics[width=.45\textwidth]{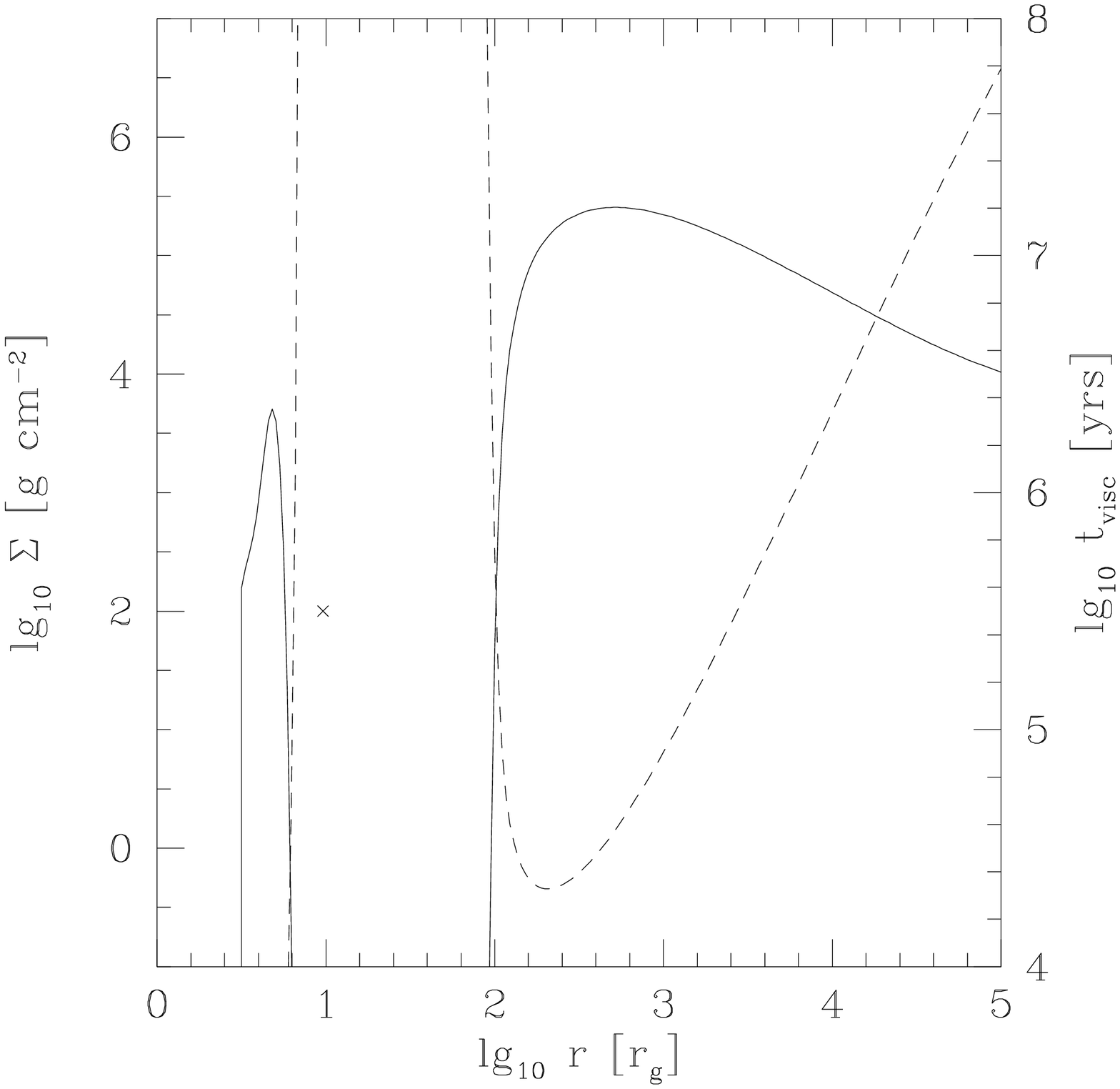}}
\end{center}
\caption{
Snapshots of the disc properties for a $10^7\Msun$ primary
  BH with a $10^6\Msun$ secondary BH (whose position is marked with an ``$\times$''), for an external accretion
  rate of $10^{-1}\Msun\,{\rm yr}^{-1}$.  
   We plot the surface 
   density $\Sigma$ (solid lines), viscous time $t_{\rm visc}$ 
   (dashed line, matched to right axis), and merger time of the 
   secondary $t_{\rm merge}$ (dotted horizontal line, match 
   to the right hand side).
  (a) shortly after initialization, when the outer disc has 
  accumulated some mass due to accretion. (b) after
  the semimajor axis has decreased by an order of magnitude due to
  tidal-viscous interaction with the outer circumbinary disc.  Note
  that the mass of the inner disc has declined significantly compared
  to (a) while the outer disc is approximately an alpha
  disc with a cutoff around the radial position of the
  secondary. (c) the semimajor axis has decreased by yet another order
  of magnitude, and the inner disc mass has reached it asymptotic
  value.  Angular momentum losses are now dominated by 
  GWs.  (d) the secondary
  tidally forces the remaining mass in the inner disc to accrete
  rapidly into the central BH. It is this final forced feeding of
  the central BH which leads to the electromagnetic precursor (see Figure \ref{fig:le_1e7}a,b).
}
\label{fig:m7mdot0.1}
\end{figure*}

\begin{figure*}
\begin{center}
\subfigure[]{ \includegraphics[width=.45\textwidth]{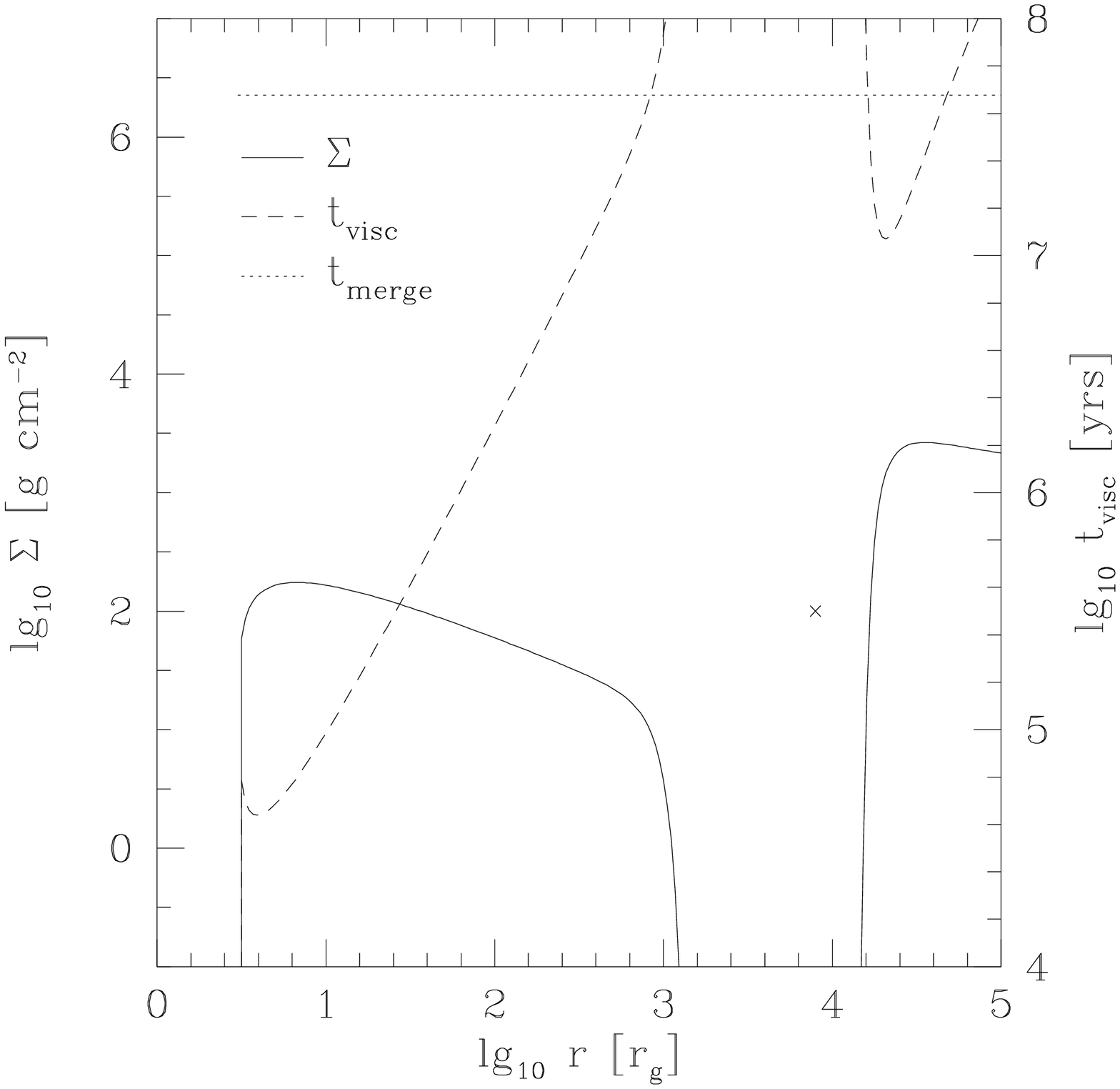}}
\subfigure[]{
  \includegraphics[width=.45\textwidth]{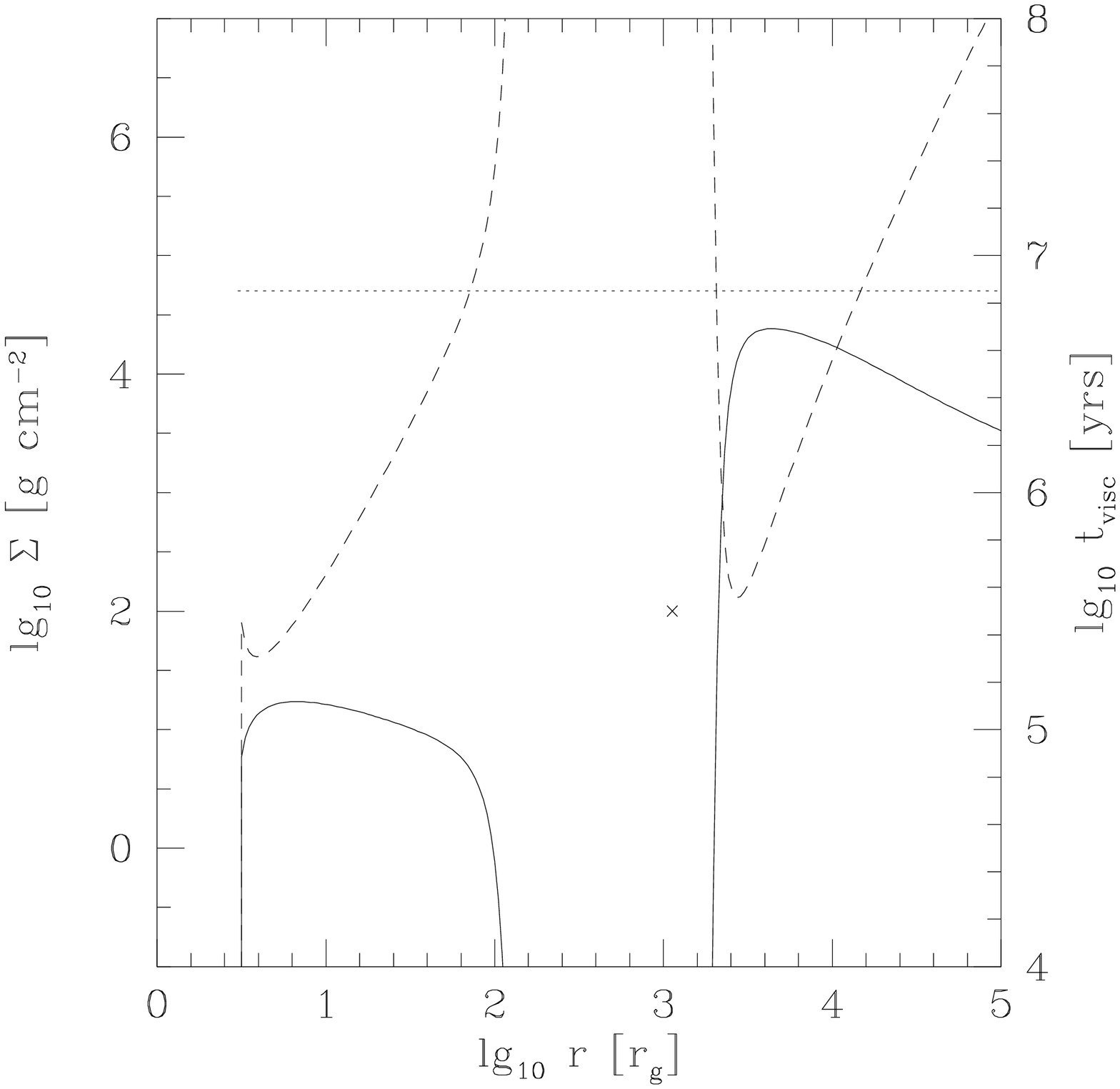}}\\ \subfigure[]{
  \includegraphics[width=.45\textwidth]{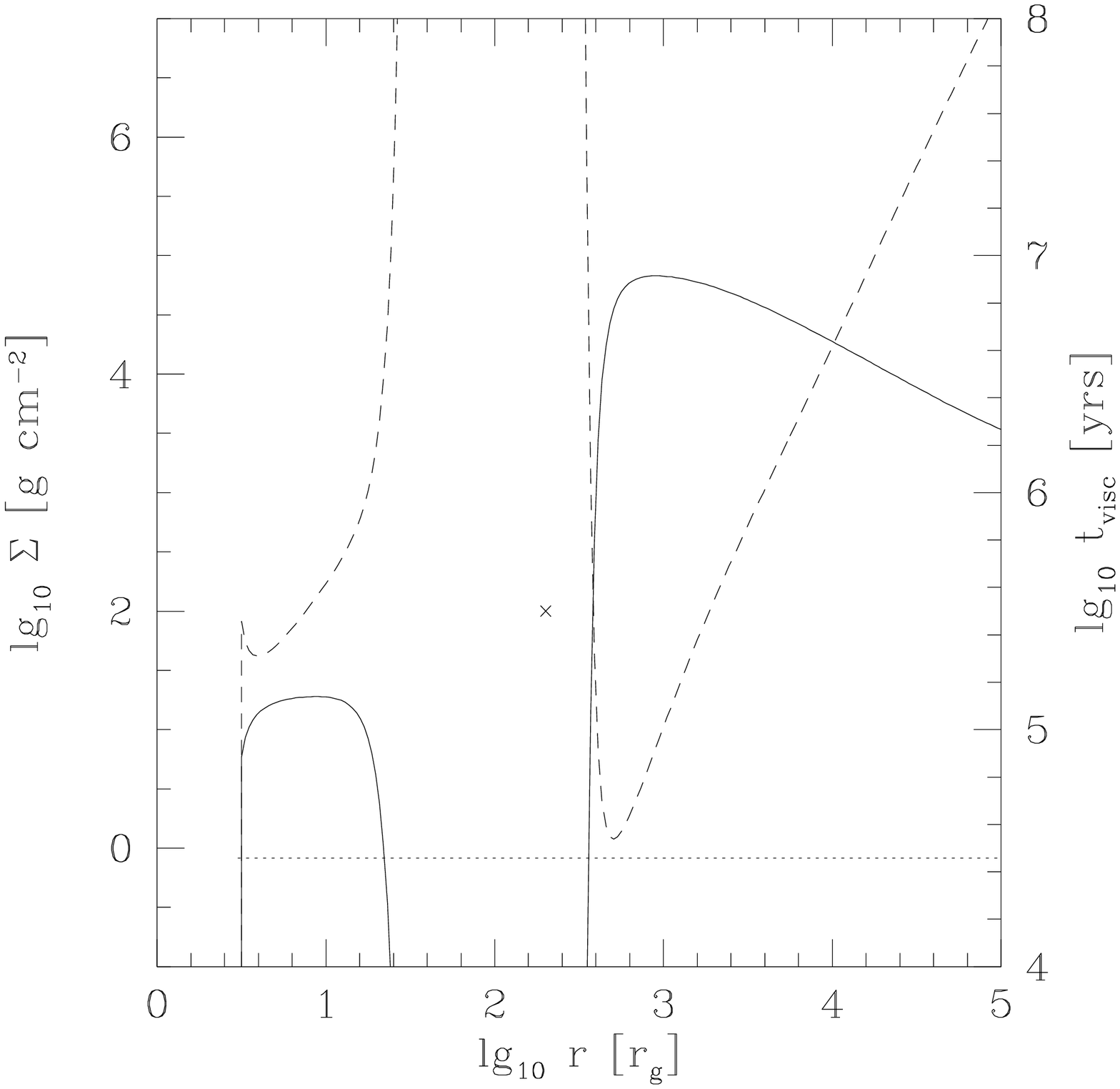}}
\subfigure[]{ \includegraphics[width=.45\textwidth]{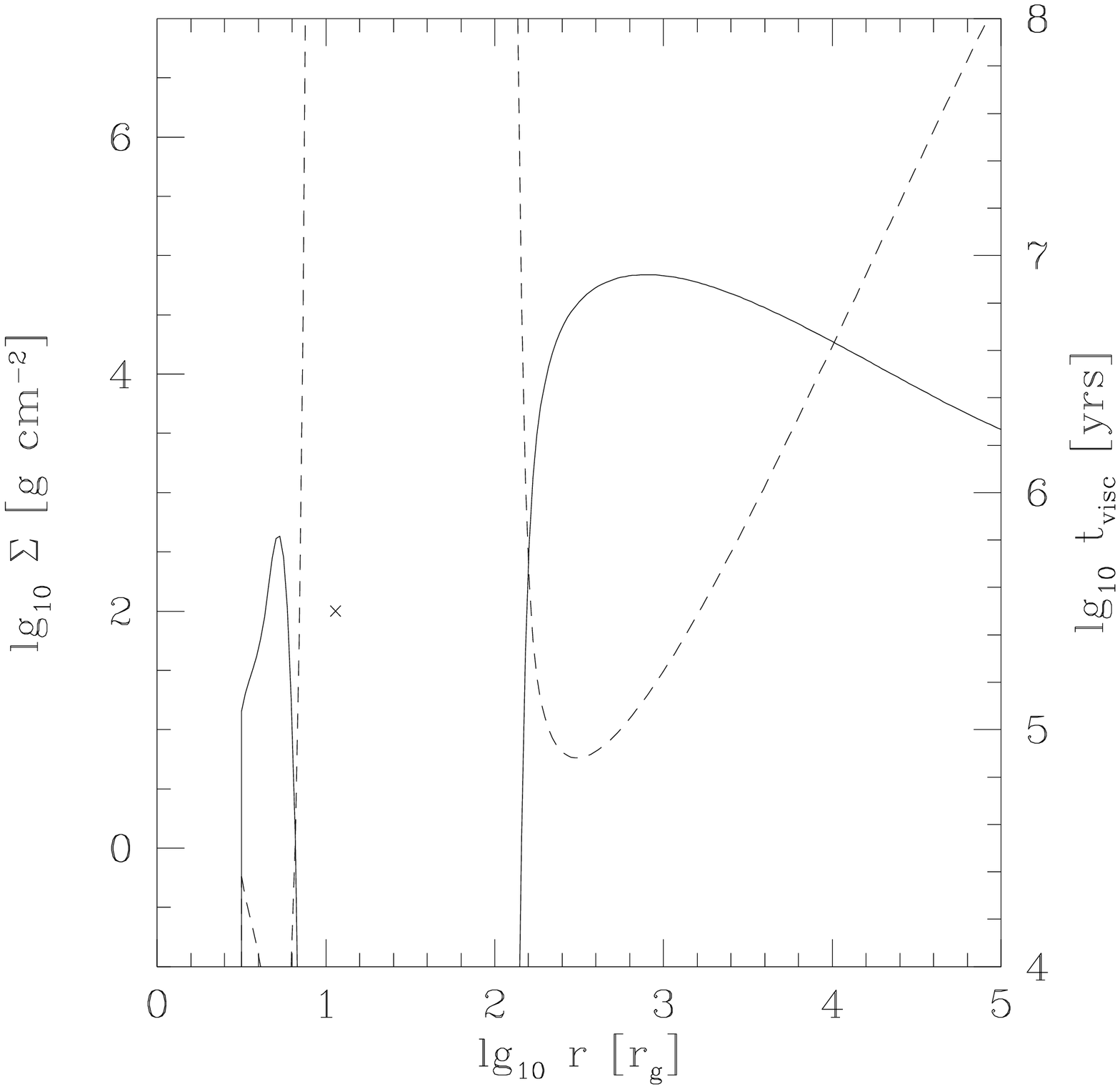}}
\end{center}
\caption{The same as Figure \ref{fig:m7mdot0.1}, but for an external
  accretion rate of $10^{-2}\Msun\,{\rm yr}^{-1}$.  The main notable
  difference is that the asymptotic mass of the inner disc is
  significantly smaller than for $\dot{M}_{\rm ext} = 10^{-1}\Msun\,{\rm
  yr}^{-1}$ case of Figure \ref{fig:m7mdot0.1}.}
\label{fig:m7mdot0.01}
\end{figure*}

We first consider the case of a $10^7\Msun$ BH with a $10^6\Msun$
secondary and an outer mass inflow rate of $\dot{M}_{\rm ext} = 0.1\Msun\,{\rm
  yr}^{-1}$ (Figure \ref{fig:m7mdot0.1}) or $\dot{M}_{\rm ext} = 0.01
\Msun\,{\rm yr}^{-1}$ (Figure \ref{fig:m7mdot0.01}).  Figure \ref{fig:m7mdot0.1} and \ref{fig:m7mdot0.01} show four different cuts from our
one-dimensional simulations to illustrate the different regimes
through which the binary and the disc evolve with time.  
In each cut, 
we plot $\Sigma$ (solid line matched to the left axis) and $t_{\rm visc}$ 
(dashed line matched to the right axis).  We also show for each cut, 
the merger time of the BBH $t_{\rm merge}$ as a horizontal dotted line 
matched to the right axis.  
In (a), the
satellite and binary have just been initialized and the outer disc has
filled due to mass inflow from the outer boundary.  In (b), the
secondary's radius has decreased by an order of magnitude due to
its tidal interaction with the viscously evolving outer disc.
Meanwhile, the inner disc has drained substantially.  In (c), the
binary has shrunk by another order of magnitude in radius due to both
tidal-viscous evolution and 
GW losses,
with the latter
becoming dominant at small radii.  Note that the inner disc density
profile has begun to invert, i.e., the surface density rises with
radius as opposed to falling.  This is due to the inability of the
inner disc to viscously respond to the increasingly rapid infall of
the secondary.  By (d), a large surface density spike has built up
near the outer edge of the inner disc.  This forcing of the inner disc
onto the primary BH causes it to brighten with a maximum luminosity, 
$L_{\rm peak}$, reached at a time, $t_{\rm peak}$, before the final merger.

A crucial aspect of the evolutionary sequence shown in Figure
\ref{fig:m7mdot0.1} and \ref{fig:m7mdot0.01} is that, during the early phases (plots a and b;
before GWs dominate), the viscous time, $t_{\rm visc} = r^2/\nu$
(dashed lines) near the outer edge of the inner disc adjusts to become
comparable to the evolving merger time, $t_{\rm merge} = |r_{\rm
  s}/v_{\rm sec}|$ (shown as a horizontal dotted line in plots a and
b).  This implies that the evolutionary time for the inner disc
is the same as the merger time of the binary.  This adjustment
occurs because a more massive disc would drain much more rapidly until
it drains until the point where its viscous time equals the 
lifetime of the system.  Similarly, a less massive disc would remain
static until the external forcing from the secondary decreases the
disc's outer radius and increases the local surface density to the
point where the local viscous time matches the merger time.  Hence the
inner disc approaches an asymptotic state that just depends on the
merger time of the binary. The binary merger time itself depends on
the external feeding rate until gravitational waves take over.
This 
self-regulated evolution explains lack of any difference of the inner disc mass on the initial
disc mass shown in Figure
\ref{fig:diff_initial}.  Note that once GWs dominate (plots c and d),
$t_{\rm visc}$ does not adjust to $t_{\rm merge}$, 
the disc can no longer respond viscously to the rapidly infalling
secondary and its dynamics becomes tidally dominated as shown vividly
in panels (d) of Figures \ref{fig:m7mdot0.1} and \ref{fig:m7mdot0.01}.
The dissipation of tidal energy and the forcing of disc material to
smaller and smaller radii is the primary means by which energy is
produced in this late stage.

As Figure \ref{fig:m7mdot0.1} and \ref{fig:m7mdot0.01} show, the general 
evolutionary sequence of a BBH system can be 
summarized as follows.  Initially the BBH starts out at large radii, where 
the tidal-viscous interaction with the outer gas disc shrinks the semi-major axis
of the BBH.  Meanwhile, the inner disc continually drains. Eventually,
the BBH shrinks until 
$r < r_{\rm GW}$, where GWs become the
principle mechanism by which angular momentum is lost.  Shortly
afterwards, the inner disc can no longer respond viscously to the
inspiralling secondary and it reaches its asymptotic mass,
$M_{\rm asym}$.\footnote{For consistency, we take the asymptotic disc
mass, $M_{\rm asym}$, to be $M_{\rm d, in}$ at $r = 100 r_g$.}  The
inner disc is forced to smaller and smaller radii by the inspiraling
secondary until it completely accretes onto the central BH.  At this
time, $t_{\rm peak}$ before the final merger, the inner disc has a 
peak luminosity, $L_{\rm peak}$.  We have run a suite of models
at different $\alpha$, $\dot{M}_{\rm ext}$, $M_{\rm BH}$, and q to explore the
parameter space of $M_{\rm d,in}$, $L_{\rm peak}$ and$t_{\rm peak}$. 
These results are summarized in Table 1 and discussed below.

\begin{table}
 \centering\label{table}
 \begin{minipage}{90mm}
  \caption{Asymptotic inner disc mass and several quantities characterizing the EM precursor for a range of black hole masses, accretion rates, and $\alpha$.
           We show $M_{\rm BH}$ [$\Msun$], q, $\dot{M}_{\rm ext}$ [$\Msun$ yr$^{-1}$], $\alpha$, $M_{\rm asym}$ [$\Msun$], 
           $t_{\rm peak}$ [days], and $L_{\rm peak}$ [ergs s$^{-1}$].}
  \begin{tabular}{rrrrrrrr}
  \hline
   $M_{\rm BH}$ &  q  & $\dot{M}_{\rm ext}$              & $\alpha$   &  $M_{\rm asym}$ & $t_{\rm peak}$ & $L_{\rm peak}$       \\
   %
\hline
   $10^6$       &     &       &       & $\times10^{-6}$  &               & $\times10^{43}$&        \\
\hline
                & 0.1 & 0.1   & 0.1   & 1.2              & 0.07          & 1.9          \\
                &     &       & 0.01  & 12               & 0.07          & 3.6          \\ 
                &     & 0.01  & 0.1   & 0.12             & 0.07          & 0.2          \\
                &     &       & 0.01  & 1.2              & 0.07          & 0.38          \\
                & 0.3 & 0.1   & 0.1   & 0.1              & 0.04          & 0.5           \\
                &     &       & 0.01  & 1                & 0.04          & 0.3           \\
                &     & 0.01  & 0.1   & 0.008            & 0.04          & 0.05          \\
                &     &       & 0.01  & 0.1              & 0.04          & 0.027          \\
\hline
   $10^7$       &     &       &       & $\times10^{-4}$ &                & $\times10^{44}$ &        \\
\hline
                & 0.1 & 0.1   & 0.1   & 6.6              & 0.7           & 2.0          \\
                &     &       & 0.01  & 460              & 0.7           & 14           \\
                &     & 0.01  & 0.1   & 0.73             & 0.7           & 0.23         \\
                &     &       & 0.01  & 6.4              & 0.7           & 2.0           \\
                & 0.3 & 0.1   & 0.1   & 0.62             & 0.4           & 0.19          \\
                &     &       & 0.01  & 5.6              & 0.4           & 1.5           \\
                &     & 0.01  & 0.1   & 0.062            & 0.4           & 0.019          \\
                &     &       & 0.01  & 0.75             & 0.4           & 0.2            \\
\hline
   $10^8$       &     &       &       & $\times10^{-2}$ &                & $\times10^{45}$ &        \\
\hline
                & 0.1 & 1     & 0.1   & 20               & 7             & 6.5           \\
                &     &       & 0.01  & 90               & 7             & 28            \\
                &     & 0.1   & 0.1   & 2.9              & 7             & 0.9           \\
                &     &       & 0.01  & 18               & 7             & 5.7           \\
                & 0.3 & 1     & 0.1   & 2.4              & 4             & 0.67          \\
                &     &       & 0.01  & 15               & 4             & 4.2           \\
                &     & 0.1   & 0.1   & 0.33             & 4             & 0.09          \\
                &     &       & 0.01  & 3.3              & 4             & 0.9           \\
\hline
\end{tabular}
\end{minipage}
\end{table}

\subsection{The Electromagnetic Precursor of the BBH Merger}\label{sec:precursor}

\begin{figure*}
\begin{center}
\subfigure[]{
\includegraphics[width=.45\textwidth]{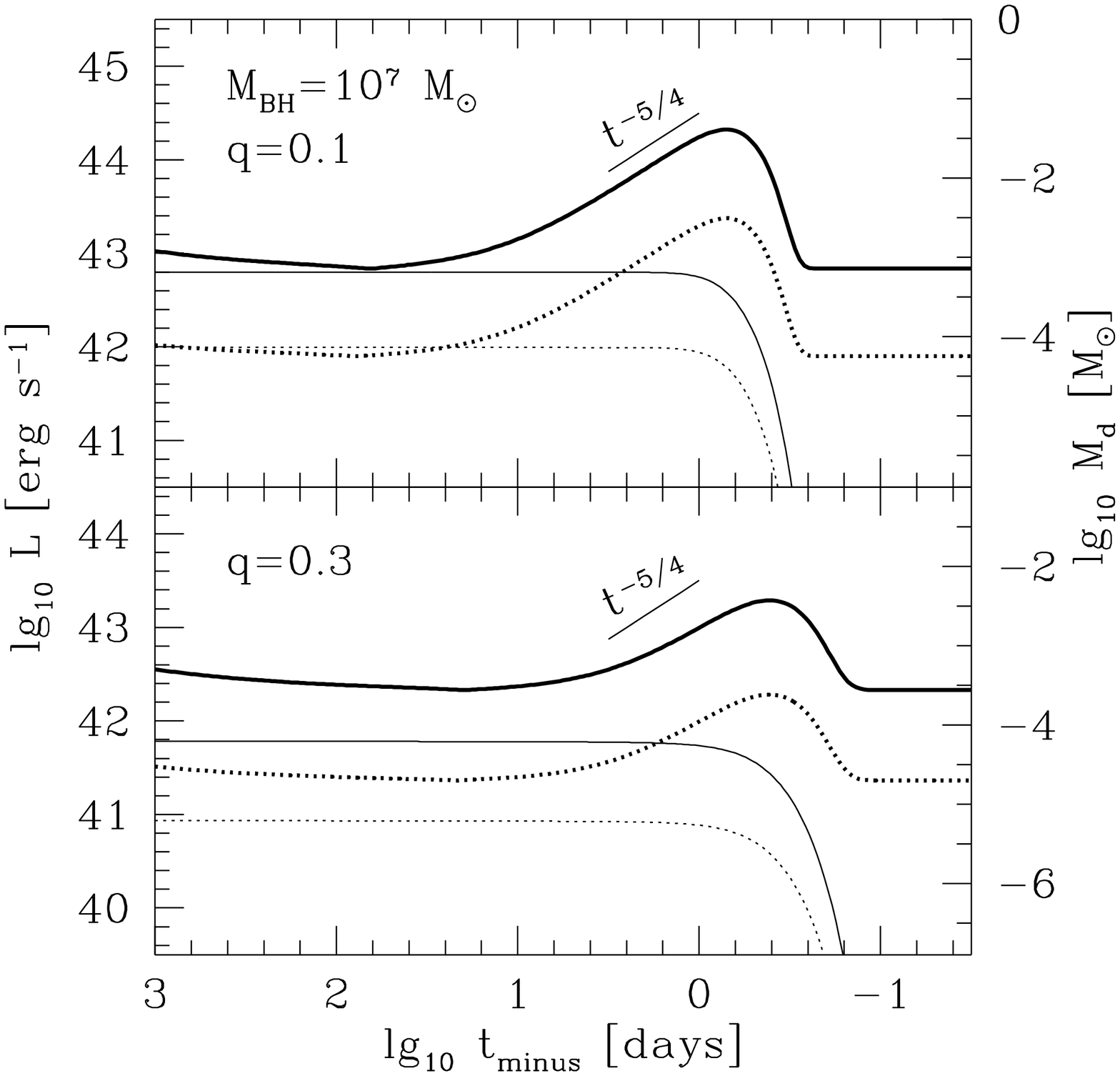}}
\subfigure[]{
\includegraphics[width=.45\textwidth]{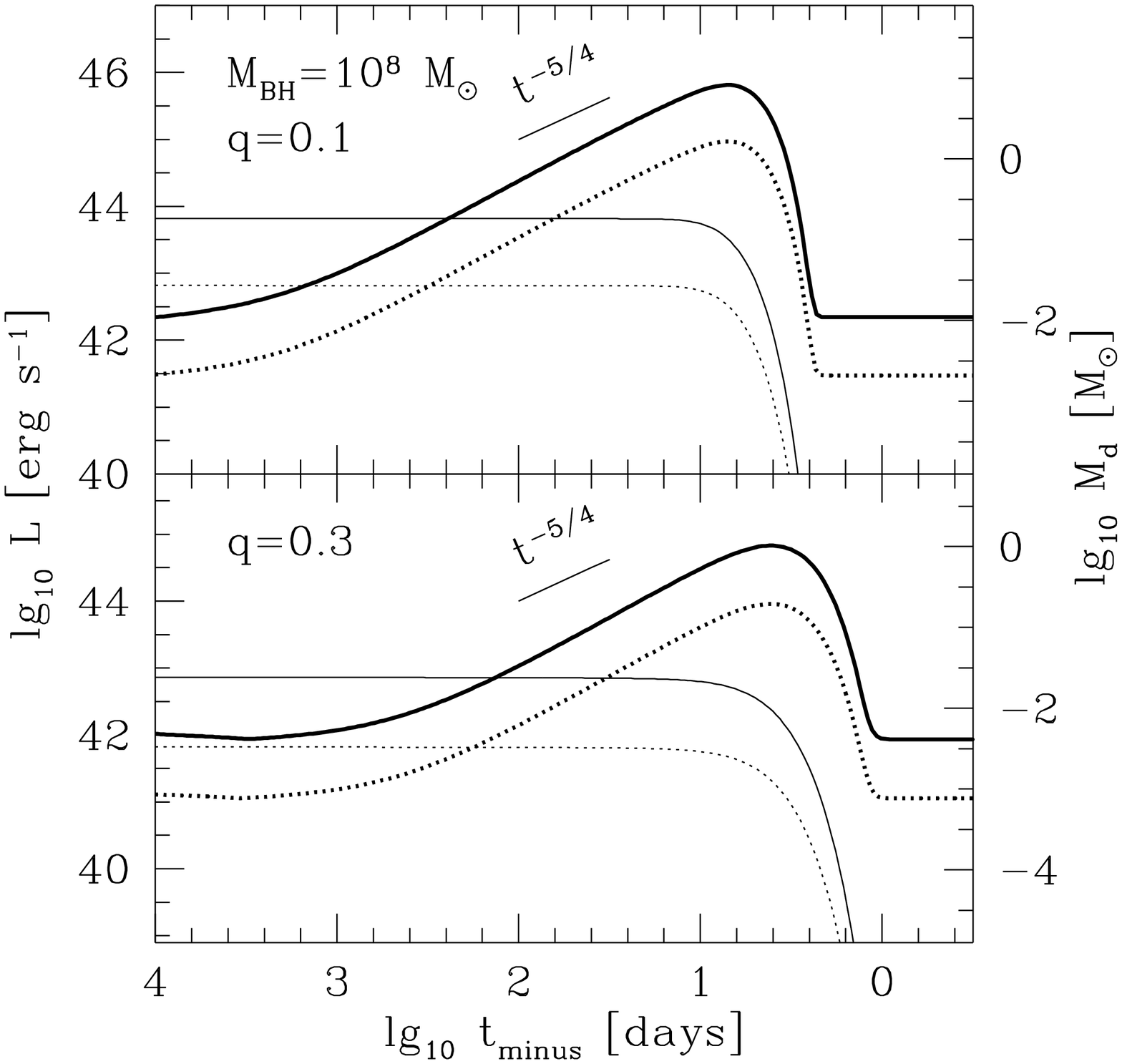}}
\end{center}
\caption{
  Bolometric luminosity (thick curves) and mass of the inner disc (thin curves)
  as a function of time before merger, $t_{\rm minus}$.  (a) Evolution 
  for a $10^{7}\Msun$ primary with a $10^6\Msun$ secondary (upper
  panel) and a $3\times 10^6\Msun$ secondary (lower panel).  We plot
  this evolution for $\dot{M}_{\rm ext}=10^{-1}\,\Msun\,{\rm yr}^{-1}$ (solid lines) and
  $\dot{M}_{\rm ext}=10^{-2}\,\Msun\,{\rm yr}^{-1}$ (dashed lines). 
  (b) Similar evolution 
  for a $10^{8}\Msun$ primary with a $10^7\Msun$ secondary (upper
  panel) and a $3\times 10^7\Msun$ secondary (lower panel) and 
  for $\dot{M}_{\rm ext}=1\,\Msun\,{\rm yr}^{-1}$ (solid lines) and $\dot{M}_{\rm ext}=10^{-1}\,\Msun\,{\rm yr}^{-1}$ (dashed lines). 
  Note that the precursor
  approximately follows a $t_{\rm minus}^{-5/4}$ power law until the peak.
%
}
\label{fig:le_1e7}
\end{figure*}

As Table 1 indicates, the peak luminosity of the electromagnetic
precursor can be appreciable even though the asymptotic mass of
the inner disc seems insignificant.  The precursor's light curve is
shown as a function of time prior to merger in Figure
\ref{fig:le_1e7}, where we plot the luminosity due to tidal and
viscous dissipation as a function of the time before merger, $t_{\rm
minus}$ (measured from when the BHs finally merge, which is not the
same as the instantaneous merger time $r_{\rm sec}/|v_{\rm sec}|$) for
the two cases of $M_{\rm BH} = 10^7\Msun$ (a) and $10^8\Msun$ (b).  The
results shown in Figure \ref{fig:le_1e7}a are for the two mass
inflow rates, $0.1\Msun\,{\rm yr}^{-1}$ (solid line) and
$0.01\Msun\,{\rm yr}^{-1}$ (dashed line), and for $q=0.1$ and $q=0.3$
(top and bottom plots respectively).  The
electromagnetic precursor manifests itself as a sudden brightening to
$\approx 0.1-1\,L_{\rm Edd}$ during the last few days before merger.
There are a few notable characteristics of the light curve.  The peak
luminosity depends only on the mass of the inner disc.  The time of
the peak $t_{\rm peak}$ depends only on the masses of the two
BHs. Finally, the light curve of the precursor has a very
characteristic power-law shape.


The first two points are clear from examining Figure
\ref{fig:le_1e7}a,b. For instance for the $M_{\rm BH} = 10^7\Msun$
($10^8\Msun$) case in Figure \ref{fig:le_1e7}a(b), the characteristic
shape and time of the peak of the light curve are the same for the
two different $\dot{M}_{\rm ext}$s, but the normalization is
different.  The normalization of the light curve depends on the mass
of the inner disc, which
in turns depends on the outer mass accretion rate and $\alpha$ (see Table 1).  
On the other hand, the
timescale of maximum luminosity only depends on the masses of
the two BHs as one can see by comparing the cases $q=0.1$
and $q=0.3$ in Figure \ref{fig:le_1e7}a,b.

The light
curve follows a characteristic power law shape, which we now derive.  
Tidal interactions between the secondary BH and the
accretion disc drive the outer edge of the disc to smaller radii and
the dissipation of the induced density waves leads to heating of the
disc.  The amount of energy dissipated at any time is
\begin{equation}\label{eq:merger luminousity}
L \approx \frac {GM_{\rm BH}M_{\rm d}}{r_{\rm sec}} t_{\rm merge}^{-1},
\end{equation}
where $M_{\rm d}$ is fixed as we are in a regime with the merging time
much shorter than the viscous time.  This mass is set primarily by the
viscosity of the disc, i.e., $\alpha$, and the external $\dot{M}_{\rm ext}$. It
directly influences the luminosity of the inner disc in this GW driven
phase of the merger.  The merging timescale due to GWs is,
\begin{equation}\label{eq:merger timescale}
t_{\rm merge} = \frac 5 8 \frac a c \frac {M_{\rm BH}} {M_{\rm sec}}
\left(\frac a {r_g}\right)^3.
\end{equation}
Combining equations (\ref{eq:merger luminousity}) and (\ref{eq:merger
  timescale}), we find 
\begin{equation}\label{eq:luminousity power law}
 L\propto t_{\rm minus}^{-5/4}.
\end{equation}
This power law shape is a good approximation to the light curve in
the more detailed calculations in Figure \ref{fig:le_1e7}a,b.

\begin{figure*}
\begin{center}
\subfigure[]{
\includegraphics[width=.45\textwidth]{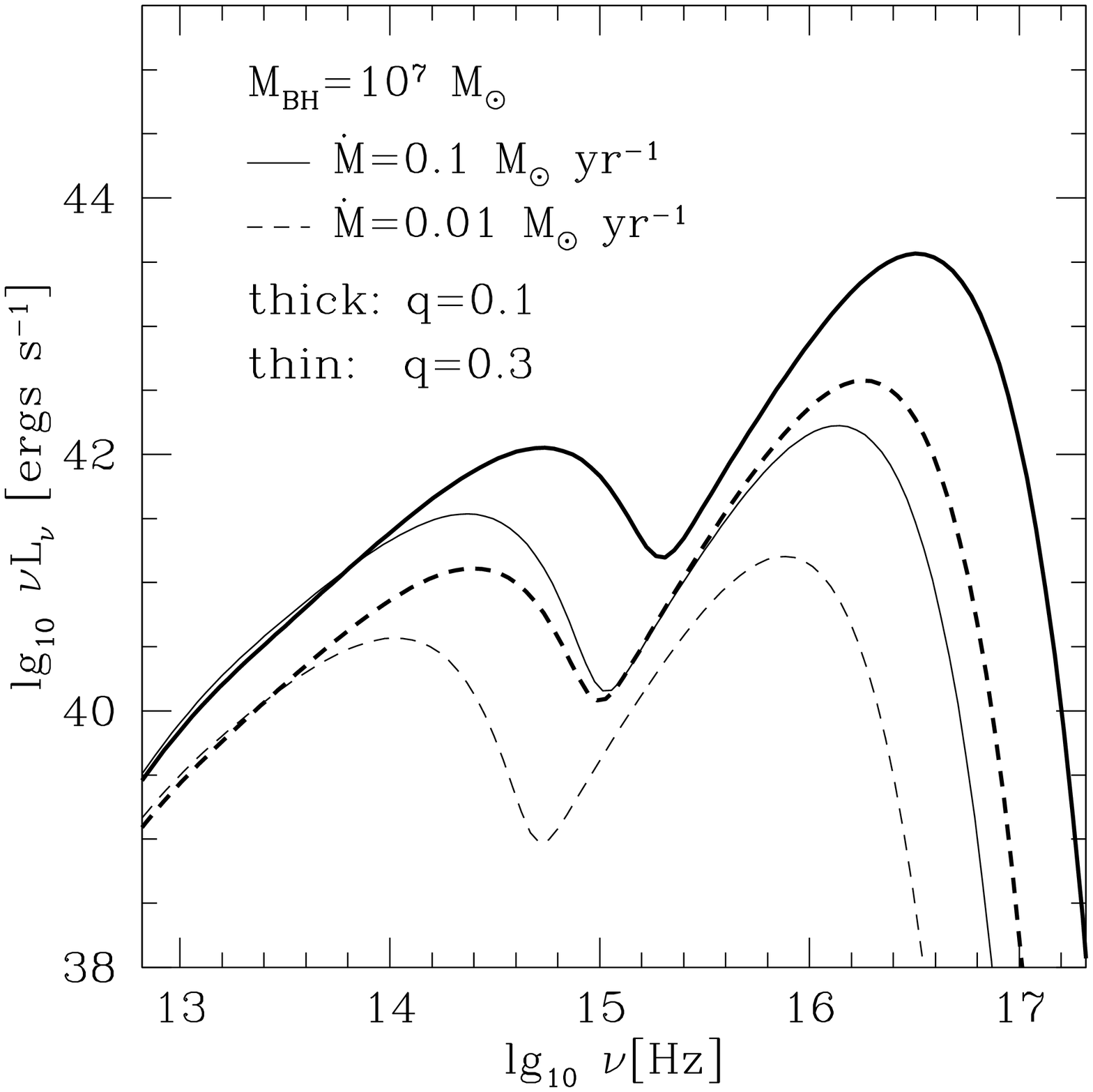}}
\subfigure[]{
\includegraphics[width=.45\textwidth]{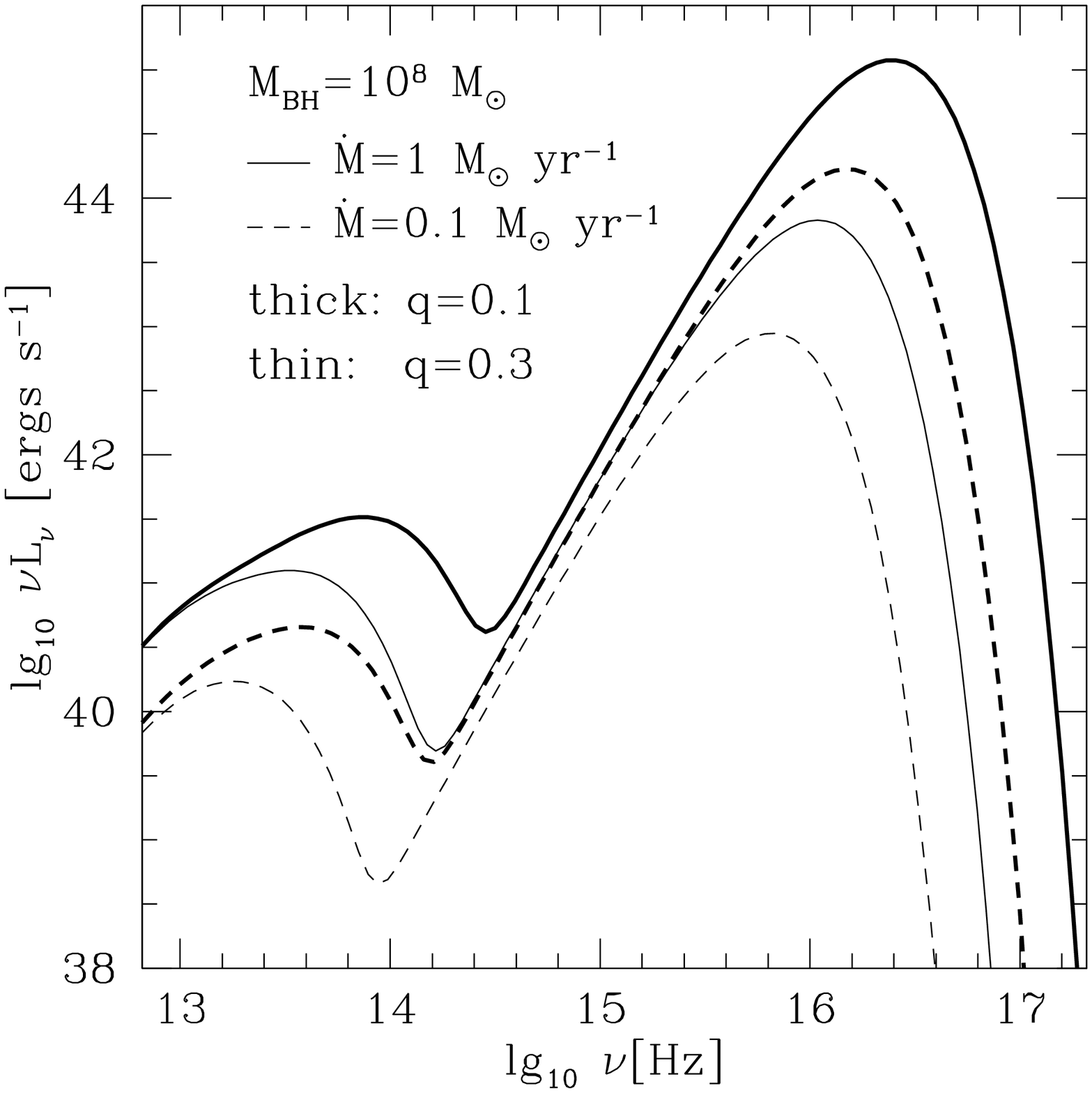}}
\end{center}
\caption{Multi-temperature blackbody spectra at $t_{\rm minus} = 1$
  day for $M_{\rm BH} = 10^7\Msun$ (left) and at $t_{\rm minus} = 10$ days for $M_{\rm BH} = 10^8\Msun$ (right) in the evolutionary scenarios of
  Figure \ref{fig:le_1e7}.  Thick lines represent a
  mass ratio $q=0.1$ and thin lines represent $q=0.3$. Note the gap in the emission between the
  low energy component, which arises from the circumbinary
  disc, and the high energy component, which arises from
  the tidally forced inner disc.}
\label{fig:spectra}
\end{figure*}

The associated effective temperature is 
$T_{\rm eff} \propto (L/r_{\rm d}(t)^2)^{1/4} \propto t_{\rm
minus}^{-7/16}$.  Figure \ref{fig:spectra} shows the corresponding multi-temperature
blackbody spectra at $t_{\rm minus} = 1$ day and $10$ days for 
$M_{\rm BH} = 10^7\Msun$ (left panel) and $10^8\Msun$ (right panel).  These spectra are
also shown for different mass ratios, $q=0.1$ (thick lines) and
$q=0.3$ (thin lines), and for different outer accretion rates,
$\dot{M}_{\rm ext} = \dot{M}_{\rm Edd}$ (solid lines) and $0.1\dot{M}_{\rm Edd}$
(dashed lines). As this plot demonstrates, the majority of the
emission prior to merger will be in the extreme UV and soft X-ray. 
As a result, the
electromagnetic precursor may be difficult to detect from the ground
or in the presence of obscuration.  We have presumed no reprocessing
of the inner disc emission by the outer disc. If present, this may increase the
luminosity of the source in more easily detectable wavelength, i.e.,
optical or near-IR, with possibly significant time delays.  
Large amplitude spiral density waves 
from the strong tidal forcing may contribute to non-thermal high energy emission.
One notable aspect of these spectra is a gap in the emission between
the low energy part, which is due to emission from the circumbinary
disc, and the high energy part, which is due to emission from the
tidally forced inner disc.

The mass in the inner disc can be
dramatically affected if additional sources of mass inflow exist beyond the viscous condition considered here.  For
instance, if the gap between the inner disc and outer disc is not
completely clear, then a small trickle of mass from the outer disc to
the inner disc could change the asymptotic inner disc mass significantly.  To
quantify this point, at the transition radius between viscous and GW
driven evolution, the mass drainage rate of the inner disc is
\begin{equation}\label{eq:mdotmax}
\dot{M} = \frac {M_{\rm in}}{t_{\rm merge}(r_{\rm GW})}\sim 10^{-10} - 10^{-9}\,\Msun\,{\rm yr}^{-1},
\end{equation}
where $t_{\rm merge} = r_{\rm sec}/v_{\rm sec}$ is the instantaneous
merger time.  This very small mass inflow rate implies that any additional
source of mass whose time averaged input rate over $t_{\rm merge} \sim
10^{6-7}\,{\rm yrs}$ 
equals or exceeds equation (\ref{eq:mdotmax}) would greatly modify the inner disc evolution.  
The calculated asymptotic disc mass may thus be considered to be a
conservative lower limit on the mass of the inner disc.

 
\section{Discussion and Conclusions}\label{sec:conclusions}

We have calculated models of BBH evolution subject to torques from a
circumbinary accretion disc and gravitational radiation losses.  In
general, the viscous time inside the orbit of the secondary BH is
much less than the timescale for tidal-viscous interaction with the
circumbinary disc to shrink the orbit of the binary.  At first glance,
this suggests that the inner disc should completely drain away.
However, we have shown that the inner accretion disc maintains a small, but
nontrivial residual mass (Fig.\ref{fig:diff_initial}), the value of which is set when GWs take
over 
(at $r_{\rm GW} \sim 500 r_{\rm g}$) as the primary mechanism
shrinking the orbit of the binary.  This inner disc, which is then
tidally forced into the primary by the secondary as it spirals inward,
brightens significantly during the last day prior to merger. The
bolometric luminosity rises as $t_{\rm minus}^{-5/4}$, peaking at one
day with a peak luminosity
of $\sim 0.1 L_{\rm Edd}$ for a representative $10^7\,M_{\odot}$ BH (see Table 1 
for results for various $\alpha$s, $M_{\rm BH}$, $\dot{M}_{\rm ext}$, and $q$)

Such an electromagnetic signature is very important for the
identification of the host galaxies of BBH mergers, but our study is
subject to several caveats.  One such caveat is related to the
persistence of a thin accretion disc at very low accretion rates.
Studies of X-ray binaries suggest that below an accretion rate
of $10^{-4} - 10^{-2} \dot{M}_{\rm Edd}$, accretion discs can
transition from a thin radiatively efficient accretion disc to a thick
radiatively-inefficient accretion flow \citep{Gallo2003,Fender2004}.  If
this were the case in our scenario prior to GW driven evolution, the
inner disc would completely drain into the primary BH and hence the
characteristic electromagnetic signature we have identified would be
absent.  We emphasize, however, that the inner disc asymptotically
approaches a gas-supported, optically-thick thin disc solution, which
in itself is stable at very low accretion rates.  Thus it is not
apriori obvious whether the transition to a geometrically thick disc
must occur in all cases.

Our calculated mass for the inner disc represents a conservative lower
limit.  When the secondary transitions from a viscosity driven
evolution to a gravitational wave driven evolution, the inner disc
mass corresponds to a mass accretion rate of $\dot{M} \sim m_{\rm
  d}/t_{\rm merge} \sim 10^{-9}\,M_{\odot}\,{\rm yr}^{-1}$.  For any
time averaged mass source of cold gas present over $t_{\rm merge}\sim
10^6$--$10^7\,{\rm yrs}$ that is greater than this, the mass of the inner disc
would be larger than what we have calculated.  For instance the mass
loss from a few massive stars such as 
found in the Galactic
center \citep{Quataert2004,Loeb2004} or an old population of stars as
in the nucleus of M31 \citep{Chang2007}, could easily enhance the mass
of the inner disc.  In addition, our one dimensional models presume
that the gap between the inner and circumbinary disc is clean and that
no mass flows between the two.  While this may not be an unreasonable
assumption given the 
comparable masses of
the two BHs, two-dimensional simulations of gaps induced by
protoplanets in a protoplanetary disc suggest that $\sim 10\%$ of
$\dot{M}_{\rm ext}$ in the outer disc can flow across the gap onto the inner
disc \citep{Lubow2006,MacFadyen2008}. Mass leakage across the gap in
our case is likely to be much smaller than what it is in the
protoplanetary case, but even a small amount of mass will make a
substantial difference to the evolution of the inner disc.

We expect the perturbation to the GW signal from the tidal forcing of
the inner disc 
to be minor 
in the sense that it would not be detectable with LISA. In the final
year of inspiral, the inner disc has reached its asymptotic mass of 
$\sim 10^{-2} - 10^{-4} \Msun$ (see Figure 5a,b). In this stage, the
inner disc modifies the torque from GWs by a factor of 
$M_{\rm d}/M_{\rm sec} \sim 10^{-10}$.  Such
a tiny mass correction is well below the best achievable precision on mass
measurements for the secondary with LISA \citep[e.g.,
][]{Hughes2002,Lang2008}.

The tidal forcing of an accretion disc by a spiraling binary is a 
somewhat unusual situation in astrophysics and the possibility that it occurs in the mergers of BBHs motivates further exploration of the
physics of this scenario.  One interesting possibility is that the
precursor re brightening event that we have identified could be
time-variable due to perturbations from the orbiting secondary.  This
is analogous to the case of superhumps, where perturbations from a low
mass secondary drives the disc in a cataclysmic variable system to be
mildly eccentric \citep{Lubow1991}. A similar
eccentricity growth has been suggested for the outer disc of BBHs by
\cite{MacFadyen2008} and \cite{Cuadra2009}.  Time variability on the
binary's period (or some multiple) would allow for an unambiguous
association of the electromagnetic signal with the GW signal measured
from future gravitational wave detectors such as LISA. The emergence
of a time variable signal from the tidally forced accretion of this fossil 
gas is worthy of further investigation.

\section*{Acknowledgments}

We thank L. Bildsten for useful discussions. P.C. thanks Kavli
Institute for Theoretical Physics for their hospitality during the
completion of this work.  This research was supported in part by the
National Science Foundation under Grant No. PHY05-51164 and by NASA
under Grant No. NNX08AH35G.  P.C. is supported in part by the Miller
Institute for Basic Research. K.M. thanks the Aspen Center for Physics
for hospitality during the completion of this work. E.Q.  is supported 
in part by NASA grant NNG06GI68G and the David and Lucile Packard Foundation.

\bibliographystyle{mn2e} 
\bibliography{harbinger}

\begin{thebibliography}{}

\bibitem[\protect\citeauthoryear{{Armitage} \& {Natarajan}}{{Armitage} \&
  {Natarajan}}{2002}]{Armitage2002}
{Armitage} P.~J.,  {Natarajan} P.,  2002, \apjl, 567, L9

\bibitem[\protect\citeauthoryear{{Armitage} \& {Natarajan}}{{Armitage} \&
  {Natarajan}}{2005}]{Armitage2005}
{Armitage} P.~J.,  {Natarajan} P.,  2005, \apj, 634, 921

\bibitem[\protect\citeauthoryear{{Arun}, {Babak}, {Berti}, {Cornish}, {Cutler},
  {Gair}, {Hughes}, {Iyer}, {Lang}, {Mandel}, {Porter}, {Sathyaprakash},
  {Sinha}, {Sintes}, {Trias}, {Van Den Broeck} \& {Volonteri}}{{Arun}
  et~al.}{2008}]{Arun2008}
{Arun} K.~G.,  {Babak} S.,  {Berti} E.,  {Cornish} N.,  {Cutler} C.,  {Gair}
  J.,  {Hughes} S.~A.,  {Iyer} B.~R.,  {Lang} R.~N.,  {Mandel} I.,  {Porter}
  E.~K.,  {Sathyaprakash} B.~S.,  {Sinha} S.,  {Sintes} A.~M.,  {Trias} M.,
  {Van Den Broeck} C.,    {Volonteri} M.,  2008, ArXiv e-prints

\bibitem[\protect\citeauthoryear{{Baker}, {Centrella}, {Choi}, {Koppitz}, {van
  Meter} \& {Miller}}{{Baker} et~al.}{2006}]{Baker2006}
{Baker} J.~G.,  {Centrella} J.,  {Choi} D.-I.,  {Koppitz} M.,  {van Meter}
  J.~R.,    {Miller} M.~C.,  2006, \apjl, 653, L93

\bibitem[\protect\citeauthoryear{{Begelman}, {Blandford} \& {Rees}}{{Begelman}
  et~al.}{1980}]{Begelman1980}
{Begelman} M.~C.,  {Blandford} R.~D.,    {Rees} M.~J.,  1980, \nat, 287, 307

\bibitem[\protect\citeauthoryear{{Bekenstein}}{{Bekenstein}}{1973}]{Bekenstein%
1973}
{Bekenstein} J.~D.,  1973, \apj, 183, 657

\bibitem[\protect\citeauthoryear{{Bode} \& {Phinney}}{{Bode} \&
  {Phinney}}{2009}]{Bode2009}
{Bode} J.~N.,  {Phinney} E.,  2009, in American Astronomical Society Meeting
  Abstracts Vol.~213 of American Astronomical Society Meeting Abstracts,
  {Observability of Circumbinary Disks Following Massive Black Hole Mergers}

\bibitem[\protect\citeauthoryear{{Bode} \& {Phinney}}{{Bode} \&
  {Phinney}}{2007}]{Bode2007}
{Bode} N.,  {Phinney} S.,  2007, APS Meeting Abstracts

\bibitem[\protect\citeauthoryear{{Boyle}, {Kesden} \& {Nissanke}}{{Boyle}
  et~al.}{2008}]{Boyle2008}
{Boyle} L.,  {Kesden} M.,    {Nissanke} S.,  2008, Physical Review Letters,
  100, 151101

\bibitem[\protect\citeauthoryear{{Chang}}{{Chang}}{2008}]{Chang2008}
{Chang} P.,  2008, \apj, 684, 236

\bibitem[\protect\citeauthoryear{{Chang}, {Murray-Clay}, {Chiang} \&
  {Quataert}}{{Chang} et~al.}{2007}]{Chang2007}
{Chang} P.,  {Murray-Clay} R.,  {Chiang} E.,    {Quataert} E.,  2007, \apj,
  668, 236

\bibitem[\protect\citeauthoryear{{Cuadra}, {Armitage}, {Alexander} \&
  {Begelman}}{{Cuadra} et~al.}{2009}]{Cuadra2009}
{Cuadra} J.,  {Armitage} P.~J.,  {Alexander} R.~D.,    {Begelman} M.~C.,  2009,
  \mnras, 393, 1423

\bibitem[\protect\citeauthoryear{{Dotti}, {Colpi}, {Haardt} \& {Mayer}}{{Dotti}
  et~al.}{2007}]{Dotti2007}
{Dotti} M.,  {Colpi} M.,  {Haardt} F.,    {Mayer} L.,  2007, \mnras, 379, 956

\bibitem[\protect\citeauthoryear{{Favata}, {Hughes} \& {Holz}}{{Favata}
  et~al.}{2004}]{Favata2004}
{Favata} M.,  {Hughes} S.~A.,    {Holz} D.~E.,  2004, \apjl, 607, L5

\bibitem[\protect\citeauthoryear{{Fender}, {Belloni} \& {Gallo}}{{Fender}
  et~al.}{2004}]{Fender2004}
{Fender} R.~P.,  {Belloni} T.~M.,    {Gallo} E.,  2004, \mnras, 355, 1105

\bibitem[\protect\citeauthoryear{{Frank}, {King} \& {Raine}}{{Frank}
  et~al.}{2002}]{Frank2002}
{Frank} J.,  {King} A.,    {Raine} D.~J.,  2002, {Accretion Power in
  Astrophysics: Third Edition}.
Accretion Power in Astrophysics, by Juhan Frank and Andrew King and Derek
  Raine, pp.~398.~ISBN 0521620538.~Cambridge, UK: Cambridge University Press,
  February 2002.

\bibitem[\protect\citeauthoryear{{Gallo}, {Fender} \& {Pooley}}{{Gallo}
  et~al.}{2003}]{Gallo2003}
{Gallo} E.,  {Fender} R.~P.,    {Pooley} G.~G.,  2003, \mnras, 344, 60

\bibitem[\protect\citeauthoryear{{Goldreich} \& {Tremaine}}{{Goldreich} \&
  {Tremaine}}{1980}]{Goldreich1980}
{Goldreich} P.,  {Tremaine} S.,  1980, \apj, 241, 425

\bibitem[\protect\citeauthoryear{{Gonz{\'a}lez}, {Hannam}, {Sperhake},
  {Br{\"u}gmann} \& {Husa}}{{Gonz{\'a}lez} et~al.}{2007}]{Gonzalez2007}
{Gonz{\'a}lez} J.~A.,  {Hannam} M.,  {Sperhake} U.,  {Br{\"u}gmann} B.,
  {Husa} S.,  2007, Physical Review Letters, 98, 231101

\bibitem[\protect\citeauthoryear{{Haiman}, {Kocsis} \& {Menou}}{{Haiman}
  et~al.}{2008}]{Haiman2008}
{Haiman} Z.,  {Kocsis} B.,    {Menou} K.,  2008, ArXiv e-prints

\bibitem[\protect\citeauthoryear{{Herrmann}, {Hinder}, {Shoemaker}, {Laguna} \&
  {Matzner}}{{Herrmann} et~al.}{2007}]{Herrmann2007}
{Herrmann} F.,  {Hinder} I.,  {Shoemaker} D.,  {Laguna} P.,    {Matzner} R.~A.,
   2007, \apj, 661, 430

\bibitem[\protect\citeauthoryear{{Hirose}, {Krolik} \& {Blaes}}{{Hirose}
  et~al.}{2009}]{Hirose2009}
{Hirose} S.,  {Krolik} J.~H.,    {Blaes} O.,  2009, \apj, 691, 16

\bibitem[\protect\citeauthoryear{{Holz} \& {Hughes}}{{Holz} \&
  {Hughes}}{2005}]{Holz2005}
{Holz} D.~E.,  {Hughes} S.~A.,  2005, \apj, 629, 15

\bibitem[\protect\citeauthoryear{{Hourigan} \& {Ward}}{{Hourigan} \&
  {Ward}}{1984}]{Hourigan1984}
{Hourigan} K.,  {Ward} W.~R.,  1984, Icarus, 60, 29

\bibitem[\protect\citeauthoryear{{Hughes}}{{Hughes}}{2002}]{Hughes2002}
{Hughes} S.~A.,  2002, \prd, 66, 102001

\bibitem[\protect\citeauthoryear{{Ivanov}, {Papaloizou} \& {Polnarev}}{{Ivanov}
  et~al.}{1999}]{Ivanov1999}
{Ivanov} P.~B.,  {Papaloizou} J.~C.~B.,    {Polnarev} A.~G.,  1999, \mnras,
  307, 79

\bibitem[\protect\citeauthoryear{{Kocsis}, {Frei}, {Haiman} \&
  {Menou}}{{Kocsis} et~al.}{2006}]{Kocsis2006}
{Kocsis} B.,  {Frei} Z.,  {Haiman} Z.,    {Menou} K.,  2006, \apj, 637, 27

\bibitem[\protect\citeauthoryear{{Kocsis}, {Haiman} \& {Menou}}{{Kocsis}
  et~al.}{2008}]{Kocsis2008a}
{Kocsis} B.,  {Haiman} Z.,    {Menou} K.,  2008, \apj, 684, 870

\bibitem[\protect\citeauthoryear{{Kocsis}, {Haiman}, {Menou} \&
  {Frei}}{{Kocsis} et~al.}{2007}]{Kocsis2007}
{Kocsis} B.,  {Haiman} Z.,  {Menou} K.,    {Frei} Z.,  2007, \prd, 76, 022003

\bibitem[\protect\citeauthoryear{{Kocsis} \& {Loeb}}{{Kocsis} \&
  {Loeb}}{2008}]{Kocsis2008b}
{Kocsis} B.,  {Loeb} A.,  2008, Physical Review Letters, 101, 041101

\bibitem[\protect\citeauthoryear{{Lang} \& {Hughes}}{{Lang} \&
  {Hughes}}{2008}]{Lang2008}
{Lang} R.~N.,  {Hughes} S.~A.,  2008, \apj, 677, 1184

\bibitem[\protect\citeauthoryear{{Lightman} \& {Eardley}}{{Lightman} \&
  {Eardley}}{1974}]{Lightman1974}
{Lightman} A.~P.,  {Eardley} D.~M.,  1974, \apjl, 187, L1+

\bibitem[\protect\citeauthoryear{{Lin} \& {Papaloizou}}{{Lin} \&
  {Papaloizou}}{1979a}]{Lin1979b}
{Lin} D.~N.~C.,  {Papaloizou} J.,  1979a, \mnras, 188, 191

\bibitem[\protect\citeauthoryear{{Lin} \& {Papaloizou}}{{Lin} \&
  {Papaloizou}}{1979b}]{Lin1979a}
{Lin} D.~N.~C.,  {Papaloizou} J.,  1979b, \mnras, 186, 799

\bibitem[\protect\citeauthoryear{{Lin} \& {Papaloizou}}{{Lin} \&
  {Papaloizou}}{1986}]{Lin1986}
{Lin} D.~N.~C.,  {Papaloizou} J.,  1986, \apj, 307, 395

\bibitem[\protect\citeauthoryear{{Lippai}, {Frei} \& {Haiman}}{{Lippai}
  et~al.}{2008}]{Lippai2008}
{Lippai} Z.,  {Frei} Z.,    {Haiman} Z.,  2008, \apjl, 676, L5

\bibitem[\protect\citeauthoryear{{Loeb}}{{Loeb}}{2004}]{Loeb2004}
{Loeb} A.,  2004, \mnras, 350, 725

\bibitem[\protect\citeauthoryear{{Loeb}}{{Loeb}}{2007}]{Loeb2007}
{Loeb} A.,  2007, Physical Review Letters, 99, 041103

\bibitem[\protect\citeauthoryear{{Lubow}}{{Lubow}}{1991}]{Lubow1991}
{Lubow} S.~H.,  1991, \apj, 381, 259

\bibitem[\protect\citeauthoryear{{Lubow} \& {D'Angelo}}{{Lubow} \&
  {D'Angelo}}{2006}]{Lubow2006}
{Lubow} S.~H.,  {D'Angelo} G.,  2006, \apj, 641, 526

\bibitem[\protect\citeauthoryear{{MacFadyen} \&
  {Milosavljevi{\'c}}}{{MacFadyen} \&
  {Milosavljevi{\'c}}}{2008}]{MacFadyen2008}
{MacFadyen} A.~I.,  {Milosavljevi{\'c}} M.,  2008, \apj, 672, 83

\bibitem[\protect\citeauthoryear{{Madau} \& {Quataert}}{{Madau} \&
  {Quataert}}{2004}]{Madau2004}
{Madau} P.,  {Quataert} E.,  2004, \apjl, 606, L17

\bibitem[\protect\citeauthoryear{{Magorrian}, {Tremaine}, {Richstone},
  {Bender}, {Bower}, {Dressler}, {Faber}, {Gebhardt}, {Green}, {Grillmair},
  {Kormendy} \& {Lauer}}{{Magorrian} et~al.}{1998}]{Magorrian1998}
{Magorrian} J.,  {Tremaine} S.,  {Richstone} D.,  {Bender} R.,  {Bower} G.,
  {Dressler} A.,  {Faber} S.~M.,  {Gebhardt} K.,  {Green} R.,  {Grillmair} C.,
  {Kormendy} J.,    {Lauer} T.,  1998, \aj, 115, 2285

\bibitem[\protect\citeauthoryear{{Milosavljevi{\'c}} \&
  {Phinney}}{{Milosavljevi{\'c}} \& {Phinney}}{2005}]{Milos2005}
{Milosavljevi{\'c}} M.,  {Phinney} E.~S.,  2005, \apjl, 622, L93

\bibitem[\protect\citeauthoryear{{O'Neill}, {Miller}, {Bogdanovic}, {Reynolds}
  \& {Schnittman}}{{O'Neill} et~al.}{2008}]{Oneill2008}
{O'Neill} S.~M.,  {Miller} M.~C.,  {Bogdanovic} T.,  {Reynolds} C.~S.,
  {Schnittman} J.,  2008, ArXiv e-prints

\bibitem[\protect\citeauthoryear{{Peters}}{{Peters}}{1964}]{Peters1964}
{Peters} P.~C.,  1964, Physical Review, 136, 1224

\bibitem[\protect\citeauthoryear{{Piran}}{{Piran}}{1978}]{Piran1978}
{Piran} T.,  1978, \apj, 221, 652

\bibitem[\protect\citeauthoryear{{Press}, {Teukolsky}, {Vettering} \&
  {Flannery}}{{Press} et~al.}{1992}]{Press1992}
{Press} W.~H.,  {Teukolsky} S.~A.,  {Vettering} W.~T.,    {Flannery} B.~P.,
  1992, {Numerical Recipes}.
Cambridge Univ. Press, Cambridge

\bibitem[\protect\citeauthoryear{{Pringle}}{{Pringle}}{1981}]{Pringle1981}
{Pringle} J.~E.,  1981, \araa, 19, 137

\bibitem[\protect\citeauthoryear{{Quataert}}{{Quataert}}{2004}]{Quataert2004}
{Quataert} E.,  2004, \apj, 613, 322

\bibitem[\protect\citeauthoryear{{Rafikov}}{{Rafikov}}{2002}]{Rafikov2002}
{Rafikov} R.~R.,  2002, \apj, 572, 566

\bibitem[\protect\citeauthoryear{{Schnittman} \& {Krolik}}{{Schnittman} \&
  {Krolik}}{2008}]{Schnittman2008}
{Schnittman} J.~D.,  {Krolik} J.~H.,  2008, \apj, 684, 835

\bibitem[\protect\citeauthoryear{{Schutz}}{{Schutz}}{1986}]{Schutz1986}
{Schutz} B.~F.,  1986, \nat, 323, 310

\bibitem[\protect\citeauthoryear{{Shakura} \& {Syunyaev}}{{Shakura} \&
  {Syunyaev}}{1973}]{Shakura1973}
{Shakura} N.~I.,  {Syunyaev} R.~A.,  1973, \aap, 24, 337

\bibitem[\protect\citeauthoryear{{Shields} \& {Bonning}}{{Shields} \&
  {Bonning}}{2008}]{Shields2008}
{Shields} G.~A.,  {Bonning} E.~W.,  2008, \apj, 682, 758

\bibitem[\protect\citeauthoryear{{Tremaine}, {Gebhardt}, {Bender}, {Bower},
  {Dressler}, {Faber}, {Filippenko}, {Green}, {Grillmair}, {Ho}, {Kormendy},
  {Lauer}, {Magorrian}, {Pinkney} \& {Richstone}}{{Tremaine}
  et~al.}{2002}]{Tremaine2002}
{Tremaine} S.,  {Gebhardt} K.,  {Bender} R.,  {Bower} G.,  {Dressler} A.,
  {Faber} S.~M.,  {Filippenko} A.~V.,  {Green} R.,  {Grillmair} C.,  {Ho}
  L.~C.,  {Kormendy} J.,  {Lauer} T.~R.,  {Magorrian} J.,  {Pinkney} J.,
  {Richstone} D.,  2002, \apj, 574, 740

\bibitem[\protect\citeauthoryear{{van de Ven} \& {Chang}}{{van de Ven} \&
  {Chang}}{2008}]{VanDeVen2008}
{van de Ven} G.,  {Chang} P.,  2008, ApJ in press, ArXiv e-prints, 0807.2437

\bibitem[\protect\citeauthoryear{{Ward}}{{Ward}}{1997}]{Ward1997}
{Ward} W.~R.,  1997, Icarus, 126, 261

\bibitem[\protect\citeauthoryear{{Ward} \& {Hourigan}}{{Ward} \&
  {Hourigan}}{1989}]{Ward1989}
{Ward} W.~R.,  {Hourigan} K.,  1989, \apj, 347, 490

\end{thebibliography}

\end{document}